\documentclass{aa}
\usepackage{txfonts}
\usepackage{graphicx}
\begin{document}

\title{The stellar populations of the bluest low surface brightness galaxies \thanks{Based on observations made with ESO Telescopes at Paranal under programme 69.B-0716.} \fnmsep \thanks{Based on observations made with the Nordic Optical Telescope, operated on the island of La Palma jointly by Denmark, Finland, Iceland, Norway, and Sweden, in the Spanish Observatorio del Roque de los Muchachos of the Instituto de Astrofisica de Canarias.}}

\author{Erik Zackrisson \inst{1}
\and
Nils Bergvall\inst{1}
\and
G\"oran \"Ostlin\inst{2}}
\offprints{Erik Zackrisson, \email{ez@astro.uu.se}}

\institute{Department of Astronomy and Space Physics, Box 515, 751 20 Uppsala
Sweden \\
\email{ez@astro.uu.se}, \email{nils.bergvall@astro.uu.se}
\and
Stockholm Observatory, SCFAB, 106 91 Stockholm, Sweden \\
\email{ostlin@astro.su.se}}

\date{Received 0000 / Accepted 0000}

\abstract{Using optical/near-IR broadband photometry together with H$\alpha$ emission line data, we attempt to constrain the star formation histories, ages, total stellar masses and stellar mass-to-light ratios for a sample of extremely blue low surface brightness galaxies. We find that, under standard assumptions about the stellar initial mass function, the H$\alpha$ equivalent widths of these objects appear inconsistent with recently suggested scenarios including constant or increasing star formation rates over cosmological time scales. In a critical assessment of the prospects of obtaining ages from integrated broadband photometry, we conclude that even with near-IR data, the ages are poorly constrained and that current observations cannot rule out the possibility that these objects formed as recently as 1--2 Gyr ago. Methods which could potentially improve the age estimates are discussed. The stellar masses of these galaxies are inferred to lie below $10^{10} \ M_\odot$. This, in combination with low ages, could constitute a problem for current hierarchical models of galaxy formation, which predict objects of this mass to form predominantly early in the history of the universe. The possibility to use the ages of the bluest low surface brightness galaxies as a test of such models is demonstrated.

}
\authorrunning{Zackrisson, E., et al.}
\titlerunning{The bluest low surface brightness galaxies} 
\maketitle

\section{Introduction}
Low surface brightness galaxies (LSBGs), here characterized by central surface brightness levels fainter than $\mu_{B,0}\sim 23$ mag arcsec$^{-2}$, have stellar populations ranging from very red ($B-V \sim 1.2$ mag) to very blue ($B-V \sim 0.2$ mag)  (e.g. O'Neil et al. \cite{oneil}). The blue LSBGs are among the most gas-rich galaxies in the local universe (e.g. McGaugh \& de Blok \cite{McGaugh & de Blok}) and have low metallicities (e.g. McGaugh \cite{mcgaugh}; R\"onnback and Bergvall \cite{ronnback2}), typically 5--20\% of the solar abundances. These properties could suggest that such galaxies, of which most have disk-like structures, may have formed recently. 

According to currently popular models of structure formation based on the cold dark matter (CDM) paradigm, low-mass galaxies should predominantly form early, whereas higher-mass objects are predicted to form through mergers and accretion at later epochs. This hierarchical scenario for galaxy formation has recently been questioned by the discovery of anti-hierarchical relations between mass and mean stellar age for low-redshift galaxies (e.g. Thomas et al. \cite{Thomas et al.}; Jimenez et al. \cite{jimenez3}) and the discovery of large populations of high-mass objects present already at high redshift (e.g. Somerville \cite{Somerville}). If the existence of truly young, low-mass galaxies could be demonstrated, this could turn out to be equally problematic. In this context, discussions on individual candidates for galaxies having formed recently, such as SBS 0335-052, Tololo 65 and IZw18 (e.g. Izotov \& Thuan \cite{Izotov & Thuan}; Kunth \& \"Ostlin \cite{Kunth & Östlin}), are however of limited value, since even in the hierarchical picture, a few stragglers are expected (e.g. Nagamine \cite{Nagamine et al.}). Only the discovery of a substantial number of young, low-mass galaxies would impose a serious constraint. Since LSBGs are believed to contribute significantly to the galaxy number density of the universe (e.g. McGaugh et al. \cite{mcgaugh2}; O'Neil \& Bothun \cite{oneil2}), and since many of these objects appear to be both unevolved and to have low masses, it becomes important to establish exactly when they formed. 

The ages of the LSBGs have long been a matter of controversy. In a search for young galaxies in the local universe, R\"onnback and Bergvall (\cite{ronnback1}) carried out one of the first multicolour studies of these objects. The sample used was selected on the basis of colour criteria, favouring extremely blue galaxies with only small radial colour gradients. R\"onnback and Bergvall ({\cite{ronnback2}) derived chemical abundances for part of this sample and found very low metallicities. This was argued to be the major reason for the blue colours, as opposed to low ages -- which was the most popular interpretation at the time. In a preliminary study (Bergvall \& R\"onnback \cite{bergvall0}), where $BVI$ photometry was interpreted using spectral evolutionary models, ages typically higher than 3 Gyr were found. Padoan et al. (\cite{padoan}) and Jimenez et al. (\cite{jimenez1}) similarly used $UBVRI$ photometry to conclude that most of the galaxies in their sample appeared to be older than about 7 Gyr. The galaxies studied were however not extremely blue, and tendencies for ages below 7 Gyr for the bluest galaxies in their sample were indeed present. Schombert et al. (\cite{schombert}) argued on the basis of their $V-I$ colours and relative HI content that the most gas-rich LSBGs should typically have mean stellar ages below 5 Gyr. Possible effects of gas infall, which could make the galaxies appear younger than their true age, were however not discussed. An opposing view was stated by van den Hoek et al. (\cite{vandenhoek}) who, based on $UBVRI$ data, concluded that the gas-rich LSBGs appeared to be very old disks with a strongly suppressed star formation activity, although the possibility that the bluest LSBGs might be dominated by stellar populations much younger than 5 Gyr could not be ruled out.

A serious shortcoming of these investigations is their reliance on optical photometry. Optical colours are, by themselves, blunt tools for constraining the ages of blue LSBGs, since the youngest stars may dominate the optical part of the spectrum. An old stellar component will thereby be extremely difficult to detect, unless strong population gradients allow the selection of regions in the galaxies where the old, red population contributes significantly to the integrated optical light. With the inclusion of near-IR broadband photometry one would naively expect to be able to constrain the ages more stringently. An investigation of this type was carried out by Bell et al. (\cite{bell}) for a sample of LSBGs covering a wide range of colours. Although a trend of lower mean stellar ages with bluer colours was found, a simplifying assumption adopted in this work was that the galaxies studied all formed 12 Gyr ago, making the average stellar age a function of metallicity and star formation history only.

Since LSBGs are believed to be more dark matter dominated then their high surface brightness counterparts (e.g. McGaugh \& de Blok \cite{McGaugh & de Blok 2}; Zavala et al. \cite{Zavala et al.}), they have become prime targets for testing predictions of dark halo density profiles through rotation curve decomposition (e.g. de Blok et al. \cite{de Blok et al.}). In this procedure, some mass-to-light ratio ($M/L$) believed to be representative of the stellar component in question is typically assumed. The most appropriate value for this parameter, which is intimately linked with both age and star formation history, is however still uncertain. It is therefore important to determine the range of $M/L$ allowed by the observed properties of the stellar populations of LSBGs.  
  
In this paper, optical/near-IR photometry and H$\alpha$ data for a sample of 9 extremely blue LSBGs are analyzed to constrain the properties of the stellar populations in these objects. Spectral evolutionary models covering a wide range of parameter values at high resolution are used to investigate their star formation histories, ages, stellar $M/L$ and total stellar masses.

\section{The data}
The optical/near-IR photometric data for the 9 blue LSBGs investigated here have  previously been presented in R\"onnback \& Bergvall (\cite{ronnback1}) and Bergvall et al. (\cite{bergvall2}). 

The targets used in the optical study of R\"onnback \& Bergvall  (\cite{ronnback1}) were selected from the ESO-Uppsala catalogue (Lauberts \& Valentijn \cite{lauberts}) with the criteria of:

1) A mean surface brightness inside the effective radius of $\mu_B^\mathrm{eff}>23.5$ mag arcsec$^{-2}$; 

2) An average colour $<$$B-R$$>$ of less than 0.5 mag in regions with surface brightness $\mu_B=20.5$--26 mag arcsec$^{-2}$; 

3) $\Delta(B-R)\le 0.2$ mag, where $\Delta(B-R)$ represents the difference in colour inside and outside the effective radius.

In the near-IR photometric study by Bergvall et al. (\cite{bergvall2}), the surface brightness criterion was however slightly relaxed. Here, galaxies with a surface brightness 0.5 mag brighter than the original study, i.e. $\mu_B^\mathrm{eff}>23.0$ mag arcsec$^{-2}$, are therefore considered.

The colour criterion targets the bluest 5\% of the LSBGs ($\mu_B^{eff}>23.5$ mag arcsec$^{-2}$) in the ESO-Uppsala catalogue. The galaxies selected this way are dust-poor (Bergvall et al. \cite{bergvall2}), which reduces the uncertainties in extinction corrections when using models to interpret their properties. The requirement of small colour gradients avoids situations where two or more separate stellar populations are located in different regions of these objects, which further simplifies the analysis. 

The data are summarized in Table \ref{observations}. The oxygen abundances, $\mathrm{O/H} = 12+\log(N(\mathrm{O})/N(\mathrm{H}))$ are from R\"onnback and Bergvall (\cite{ronnback2}). The redshift and HI mass of 074-16 has been adopted from Kilborn et al. (\cite{Kilborn et al.}). In all other cases, the HI data stem from Bergvall et al. (\cite{bergvall4}). The H$\alpha$ equivalent widths, EW(H$\alpha$), of 3 galaxies are derived from narrowband images obtained at the ESO/MPI 2.2 m telescope, La Silla, and at the NOT 2.5 m telescope on La Palma. The estimated $1\sigma$ errors are $\sigma(B-V)=0.04$ mag, $\sigma($$<$$B-R$$>$$)=0.10$ mag, $\sigma(V-i)=0.06$ mag, $\sigma(B-J)=0.08$ mag, $\sigma(J-H)=0.12$ mag, $\sigma(H-K')=0.14$ mag, $\sigma(\mathrm{O/H})=0.1$ and $\sigma(\mathrm{EW}(\mathrm{H}\alpha))=20\%$. 

In Table \ref{observations}, both the central surface brightness of the exponential disk and the ``true'' surface brightness of the centre are presented. This is particularly useful in the cases where the surface brightness is low. As was pointed out by Bergvall et al. (\cite{bergvall2}), there is a tendency of a depression relative to an exponential light distribution to develop in the centre of the galaxies when going to extremely low levels of surface brightness. Hence, the central disk surface brightness will be inadequate to describe how dim the disk really is. Such cases are e.g. \object{ESO 505-04}, \object{ESO 546-34} and \object{ESO 602-26}. The central surface brightness entries have been corrected to face-on projection for the purpose of standardization. It should be noted that these corrections may be too large for individual objects if the disk contains an embedded stellar component with a structure different from that of a thin disk, e.g. a bulge. Such effects are however negligible for the sample as a whole.

\begin{table*}
\caption[ ]{The integrated properties of our sample of 9 blue LSBGs. The morphological types have been determined by R\"onnback \& Bergvall (\cite{ronnback1}). Absolute magnitudes are based on distances derived from the model by Schechter (\cite{Schechter}), assuming a distance to the Virgo cluster of 17 Mpc and a Hubble constant of $H_0=75$ km s$^{-1}$Mpc$^{-1}$. Unless stated otherwise, $m_B$ represents the integrated $B$-band magnitude inside the $\mu_{B,0}$ = 26.5 mag arcsec$^{-2}$ isophote. Here, $\mu_0$ represents the ``true'' central surface brightness and $\mu^D_0$ the central surface brightness of the disk extrapolated to the centre. Both have been corrected for inclination assuming an infinitely thin disk. $BVR$ are in the Cousins system, $i$ in the Thuan-Gunn system and $JHK'$ in the Johnson system. All magnitudes have been corrected for galactic extinction (Burstein \& Heiles \cite{Burstein & Heiles}). EW represents the equivalent width of the H$\alpha$ emission line. All photometric data come from R\"onnback \& Bergvall (\cite{ronnback1}) and Bergvall et al. (\cite{bergvall2}), except the $<$$B-R$$>$ entries, which come from the ESO/Uppsala survey (Lauberts \& Valentijn \cite{lauberts}).}
\begin{flushleft}
\begin{tabular} {lllllllllllllll}
\noalign{\smallskip}
\hline
\noalign{\smallskip}
ESO id.  & Type &$m_\mathrm{B}$  &$M_\mathrm{B}$  & $\mu_{B,0}$ & $\mu^D_{B,0}$ & $B-V$ & $<$$B-R$$
>$ & $V-i$ & $B-J$  & $J-H$  & $J-K'$  & O/H & EW & $\log$ \\
& &  &  & (mag. & (mag. &  &  & & & & & & (\AA ) & $M(\mathrm{HI})$\\
&  &  & & arcsec$^{-2}$) & arcsec$^{-2}$)  &  &  &  &  &  &  &  &  & ($M_\odot$) \\
\noalign{\smallskip}
\hline
\noalign{\smallskip}
074-16  & SBcd & 16.0     & $-17.3$    & 23.9  & 23.6 & 0.44 & 0.42 &  0.08 & 1.76  &  0.5 &      &      &    & 9.69\\ 
084-41  & Sd   & 17.2     &            & 25.2  & 24.7 & 0.44 & 0.47 &  0.08 & 1.71  &      &      &      &    &\\
146-14  & Sdm  & 15.4$^\mathrm{a}$ & $-16.9$    & 23.6  & 23.8 & 0.36 & 0.30 &  0.02 & 1.72  &  0.7 & 0.7  & 7.61 & 37 & 9.57\\
288-48  & Pec. & 16.8     & $-14.4$    & 22.0  & 23.3 & 0.43 & 0.41 &  0.02 & 1.70  &      &      &      &    &\\
462-36  & SBd  & 16.4     &            & 22.8  & 23.1 & 0.46 & 0.42 &  0.10 & 1.94  &      & 0.7  &      &    &\\
505-04  & Sm   & 15.3     & $-16.9$    & 22.7  & 21.3 & 0.49 & 0.42 & $-0.01$ &       &      &      & 7.68 & 15 & 8.67\\
546-09  & Sm   & 15.2     & $-18.0$    & 21.9  & 22.7 & 0.43 & 0.43 &  0.24 & 1.80  &      &      & 7.95 &    & 9.27\\
546-34  & Sm   & 15.7     & $-15.8$    & 24.5  & 22.8 & 0.38 & 0.43 &  0.01 & 1.40  &      &      & 7.64 & 37 & 8.81\\
602-26  & Pec. & 17.1     &            & 24.7  & 23.6 & 0.31 & 0.39 &  0.05 & 1.40  & 0.7  &      &      &    &\\
\noalign{\smallskip}
\hline
\end{tabular} \\
a) Magnitude based on the $\mu_B$ = 25 mag arcsec$^{-2}$ isophote \\
\end{flushleft}
\label{observations}
\end{table*}

\section{Spectral evolutionary modelling}
When analyzing the integrated properties of galaxies, it is important to use spectral evolutionary models as closely adapted to the known conditions of the target objects as possible. Here, the model of Zackrisson et al. (\cite{Zackrisson}, hereafter Z01) is mainly used, with the publicly available model P\'EGASE.2 (Fioc \& Rocca-Volmerange \cite{Fioc & Rocca-Volmerange}) employed to a lesser extent to test the robustness of all important conclusions. These models are adopted because of their capability to predict the spectral energy distribution of stellar populations with ages spanning from young to very old, as well as the corresponding component of nebular emission, which can be crucial for the interpretation of optical and near-IR photometry in star-forming galaxies  (see e.g. Z01, \"Ostlin et al. \cite{Östlin et al.}).

Z01 is based on the method of isomass synthesis and employs synthetic stellar atmospheres by  Lejeune (\cite{Lejeune et al.}) and Clegg 
\& Middlemass (\cite{Clegg & Middlemass}), together with stellar evolutionary tracks mainly from the Geneva group. The stellar component includes
pre-main sequence evolution and a stochastic treatment of horizontal branch morphologies at low metallicities. Gas continuum and emission lines are predicted using the photoionization code Cloudy version 90.05 (Ferland \cite{Ferland}). For each time step, the spectral energy distribution of the integrated stellar population is used as input to Cloudy, which makes the computation time-consuming, but provides realistic temporal evolution of the nebular component.

P\'EGASE.2 is based on the method of isochrone synthesis, employs the same synthetic stellar atmospheres as Z01, but stellar evolutionary tracks mainly from the Padova group. Using a simple prescription for the nebular component, it predicts both continuum and line emission of the ionized gas, although with non-evolving emission line ratios. 

\subsection{Extinction corrections}
We apply corrections for intrinsic extinction by assuming an extinction curve characteristic of the Milky Way (Miller \& Mathews \cite{Miller & Mathews}). Since the extinction, based on the measured Balmer decrement (R\"onnback \cite{Rönnback}), appears to be very low in these objects, we consider $c=\log_{10}(F_{\mathrm{H}\beta,0}/F_{\mathrm{H}\beta})=0.15\pm 0.05$ (corresponding to $E(B-V)=0.03$--0.07 mag), where $F_{\mathrm{H}\beta}$ represents the observed H$\beta$ emission line flux and $F_{\mathrm{H}\beta,0}$ the H$\beta$ emission line flux in the absence of extinction. 

\subsection{Chemical abundances}
Because of the very low metallicities of blue LSBGs (5--20\% solar) and the small metallicity dispersion among their different HII-regions (R\"onnback \cite{Rönnback}), constant metallicities of $Z=0.001$ and $Z=0.004$ are adopted for the grid of evolutionary sequences used. A more realistic approach would be to use a self-consistent evolution of the metallicity, and allow the rates of infall of primordial gas or blow-out of metal-enriched material to be free parameters. The measured metallicities for each galaxy could then be used to determine the quality of a given fit. This would not give rise to a constant metallicity, but rather a metallicity distribution with a luminosity-weighted average corresponding to the measured value. Such an approach would however increase the required grid of model evolutionary sequences by a large factor and is not essential here. 

\subsection{The initial mass function}
The initial mass function (IMF) of the stellar component is assumed to follow a single-valued power law
\begin{equation}
\Phi(m,\alpha)=\frac{\mathrm{d}N}{\mathrm{d}m}\propto m^{-\alpha}
\end {equation}
inside the mass interval [$M_\mathrm{low}$, $M_\mathrm{up}$], where a slope of $\alpha=2.35$ corresponds to the commonly adopted Salpeter IMF. The Salpeter slope is often assumed to provide an adequate description of stars more massive than $\approx 0.5 \ M_\odot$ for both the Galaxy and the Large Magellanic Cloud, but has been shown to significantly overpredict the number of stars less massive than this (e.g. Kroupa \cite{Kroupa}). Although the necessary modification of the IMF at the lowest masses has a pronounced effect on the mass budget of normal stellar populations, the affected low-mass stars contribute very little to the integrated light. A common strategy is therefore to use a single-valued power law when calculating the spectral energy distribution of synthetic stellar populations, and to rescale the mass-dependent predicted quantities (e.g. stellar $M/L$ ratios) to more realistic IMFs afterwards. This is the procedure adopted in the following.  

It should however be noted that much probably remains to be learned about the IMF and its cosmological evolution. It has been argued that the IMF could have been more top-heavy in the early universe (e.g. Larson \cite{Larson}), with a slope flatter than the Salpeter value by $\Delta \alpha \approx 0.4$. Conversely, it has also been pointed out that the local high-mass IMF may  actually be steeper than the Salpeter value by $\Delta \alpha \approx 0.4$, due to unresolved binary systems (e.g. Scalo \cite{Scalo}; Kroupa \cite{Kroupa}). To test the effects of IMF variations within this range of plausible slopes, spectral evolutionary scenarios based on single-valued IMFs with $\alpha=1.85$--2.85 are therefore used in this study.
 
Although the upper mass limit $M_\mathrm{up}$ is often assumed to lie around $120 \ M_\odot$, the precise value and its cosmic variance remain very uncertain due to small number statistics. For this reason, spectral evolutionary sequences with $M_\mathrm{up}=40 \ M_\odot$ are also tested.

\subsection{Star formation history}
Three different kinds of star formation histories (SFHs) are considered: a star formation rate (SFR) which stays constant during a given time and then drops to zero; an exponentially decreasing SFR ($\propto \exp{(-t/\tau)}$, with e-folding decay rate $\tau>0$); and an exponentially increasing SFR (same, but with $\tau<0$). The first is suitable for abrupt starbursts (with duration of bursts typically 10--100 Myr), whereas the second is more appropriate for late-type galaxies with signs of star formation over significant fractions of the age of the universe. The third has been inspired by the models of Boissier et al. (\cite{Boissier et al.}) in which low surface brightness galaxies differ from high surface brightness ones in the larger angular momenta of their dark matter halos. Their models predicts that the SFRs of the galaxies with lowest mass and highest spin should increase by several orders of magnitude over a Hubble time. Bell et al. (\cite{bell}) also find evidence that an increasing SFR is required to explain the colours of certain LSBGs. Although the SFHs predicted in Boissier et al. (\cite{Boissier et al.}) cannot be perfectly reproduced using an exponentially increasing SFR, we find that a $\tau=-2$ Gyr scenario increases the SFR by roughly the right amount over the age of the universe to accord with their steepest SFH predictions. Scenarios with a constant SFR or an exponentially increasing one with $\tau=-15$ Gyr are considered adequate to reproduce their less dramatic SFH slopes, except during the earliest phase of evolution.  

\begin{table}[t]
\caption[]{The grid of Z01 evolutionary sequences. All sequences assume $Z_\mathrm{gas}=Z_\mathrm{stars}$, a gas covering factor of 1.0, a hydrogen number density of 100 cm$^{-3}$ and the IMF slope, $\alpha$, to be valid throughout the stellar mass range 0.08--120 $M_\odot$. The grid consists of all possible combinations of the parameter values listed below. In addition to these, 108 sequences with $M_\mathrm{up}=40\ M_\odot$ were generated for test purposes with variations in IMF slope, metallicity, star formation history and redshift, but with fixed $M_\mathrm{tot}=10^9 \ M_\odot$ and filling factor 0.01.}
\begin{flushleft}
\begin{tabular}{ll} 
\hline
\hline
$M_\mathrm{tot}$ ($M_\odot$) & $10^9$, $10^{10}$ \cr
IMF, $\alpha$ & 1.85, 2.35, 2.85 \cr
SFH & c$10^8$, e$5\times10^8$, e$1\times10^9$, e$3\times10^9$, e$6\times10^9$\cr
    & e$15\times10^9$, c$15\times10^9$, e$-15\times10^9$, e$-2\times10^9$\cr
Metallicity, $Z $ & 0.001, 0.004 \cr
Filling factor & 0.01, 0.1\cr
Redshift, $z$ & 0.00, 0.01\cr
\hline
\end{tabular}\\
IMF: $dN/dm \propto m^{-\alpha}$\\
SFH=Star formation history.\\
c=Constant star formation rate during the subsequent number of years.\\
e=Exponentially declining/increasing star formation rate  with $\tau$ equal to the subsequent number of years. \\
\label{modelgrid}
\end{flushleft}
\end{table}
\begin{table}[t]
\caption[]{The grid of P\'EGASE.2 evolutionary sequences. All sequences assume the IMF slope, $\alpha$, to be valid throughout the stellar mass range 0.08--120 $M_\odot$. The grid consists of all possible combinations of the parameter values listed below. The abbreviations used are described in Table \ref{modelgrid}}
\begin{flushleft}
\begin{tabular}{ll}
\hline
\hline
IMF, $\alpha$ & 1.85, 2.35, 2.85 \cr
SFH & c$10^8$, e$5\times10^8$, e$1\times10^9$, e$3\times10^9$, e$6\times10^9$\cr
    & e$15\times10^9$, c$15\times10^9$, e$-15\times10^9$, e$-2\times10^9$\cr
Metallicity, $Z $ & 0.001, 0.004 \cr
\hline
\end{tabular}\\
\label{modelgrid2}
\end{flushleft}
\end{table}

\subsection{Other model parameters}
In Z01, the properties of the nebular component is regulated by several parameters not considered in P\'EGASE.2, some of which can have a significant effect on predicted broadband magnitudes (see Z01 for a demonstration): the metallicity of the gas, the covering factor, the filling factor, the gas density and the total mass of the stellar component. 
Here, the metallicity of the gas, $Z_\mathrm{gas}$, is assumed to be the same as that of the stellar population, $Z_\mathrm{stars}$. Since stars pollute the interstellar medium with metal-enriched material, models of chemical evolution generally predict the metallicity of the stars to be lower than that of the surrounding gas. The approximation used here is nonetheless adequate as long as the metallicity is low.  

The covering factor regulates the fraction of the stellar component covered by gas (in terms of surface area). Since blue LSBGs are gas-rich and no studies to date have discovered Lyman-continuum leakage from any LSBG, it is likely that the covering factor is very high. Here, we therefore consider a cover factor of unity only. 

The filling factor, defined as the fraction of the total volume of the Str\"omgren sphere occupied by gas, is constrained by observations to lie in the range 0.01--0.1 (Martin \cite{martin}; Miller \cite{miller}), which is the interval adopted here.

The hydrogen number density of the gas condensations inside the Str\"omgren sphere, has been taken to be 100 cm$^{-3}$, based on Copetti et al. (\cite{copetti}).

In Z01, the nebular component is computed under the assumption that the stars of a galaxy are concentrated to the centre of a single spherical nebula, rather than distributed among a large number of smaller Str\"omgren spheres. This approximation gives reasonable results for the hydrogen Balmer lines used in this study, but may not be as good for emission lines more sensitive to the detailed temperature and ionization structure of the cloud. In the framework of this approximation, the number of Lyman continuum photons striking the illuminated face of the cloud at a given population age, $t$, is regulated by SFR($t$) and the total gas mass, $M_\mathrm{tot}$, which may eventually be converted into stars during the course of a given SFH. Since previous investigations have found the baryonic mass of blue LSBGs to lie in the interval $10^9$--$10^{10} \ M_\odot$ (De Blok \& McGaugh \cite{deblok2}), this is the range adopted here. In the case of star formation rates increasing with time, $M_\mathrm{tot}$ is defined as the total gas mass that would have been consumed after 15 Gyr. It should be noted that Z01 does not assume that gas expelled from the stellar population can be recycled into subsequent generations of stars.

Finally, the conversion from synthetic population spectra to fluxes in broadband filters depends on the redshift of the target galaxies. Due to the galaxy diameter limit of the ESO-Uppsala catalogue, none of our objects are expected to lie significantly further away than $z\approx 0.01$. For this reason, only redshifts in the range $z=0$--0.01 are considered.

\subsection{Model grids} 
Using Z01, a grid of 540 evolutionary sequences has been generated with parameter values summarized in Table \ref{modelgrid}, constituting the most extensive set of spectral synthesis predictions used for the study of LSBGs so far. In addition, a smaller set of 54 evolutionary sequences, with parameter values summarized in Table \ref{modelgrid2}, was computed using P\'EGASE.2 to be used for consistency testing. Each sequence runs from ages of 1 Myr to 15 Gyr in small time steps.

\subsection{H$\alpha$ equivalent widths}
In addition to broadband fluxes, EW(H$\alpha$) are also used as a constraint when determining the properties of the stellar populations of the galaxies in our sample. Although the required H$\alpha$ narrowband data are available only for three of our nine objects (with EW(H$\alpha)=37$, 15 and 37 \AA{} respectively), the global EW(H$\alpha$) of blue LSBGs appear to typically lie inside a limited interval. Narrowband images of two additional blue LSBGs included in the original sample but lacking photometry (\object{ESO 359-31} and \object{ESO 405-06}), reveal EW(H$\alpha$) of 19 and 25 \AA{} respectively. The equivalent widths derived from VLT/FORS2 long-slit spectra along the major axis of four highly inclined LSBGs (\object{ESO 031-13}, \object{ESO 462-32}, \object{ESO 532-32}, \object{ESO 548-04}) with colours very similar to those in the present sample (Zackrisson et al., in preparation) are EW(H$\alpha)=30$, 39, 43 and 18 \AA. Due to the largely transparent disks of LSBGs (e.g. Matthews \& Wood \cite{Matthews & Wood}), these long-slit observations should provide reasonable estimates of the global EW(H$\alpha$).  With $\sigma(\mathrm{EW(H\alpha))}=20\%$, the $\mathrm{EW(H\alpha)}$ measurements for all 9 galaxies approximately lie inside the interval 10--60 \AA{} at the $2\sigma$ level, which is the adopted range of EW(H$\alpha$) considered acceptable for successful model fits in this study. 

\subsection{Homogeneity of the stellar population}
The small colour gradients in our objects indicate that their stellar populations must be fairly homogeneous. The small dispersion in EW(H$\alpha$) among blue LSBGs furthermore shows that their star formation history cannot have proceeded very stochastically, i.e. through violent starbursts, unless the IMF is peculiar. 

However, in a study of blue compact galaxies, Bergvall and \"Ostlin (\cite{bergvall3}) detected abnormally red colours, typical of a metal-rich population, in the faint halos of these objects (which are generally considered to be very metal-poor). One remote possibility is that blue LSBGs may have formed recently as a consequence of a merger between a relatively metal-rich early-type galaxy and an intergalactic gas cloud. In such a scenario, the stars formed in the merger event could very well be outshining the older population, giving the appearence of a genuinely young galaxy. The old, underlying host would then only be observable in the very faint outskirts of the blue LSBG, analogous to the case of blue compact galaxies. This scenario may however be tested by studying the halo colours of nearly edge-on galaxies, of which \object{ESO 146-14} from our sample is one example. Within the error bars, we do not detect any significant gradients in the optical/near-IR colour profiles in the direction perpendicular to the major axis of the disk. For this reason, mixed populations more extreme than those already produced by the adopted range of star formation histories, e.g. by adding a single-age young population to a single-age old one, are not considered in the present study. 
\begin{figure*}[t]
\centering
\resizebox{\hsize}{!}{\includegraphics{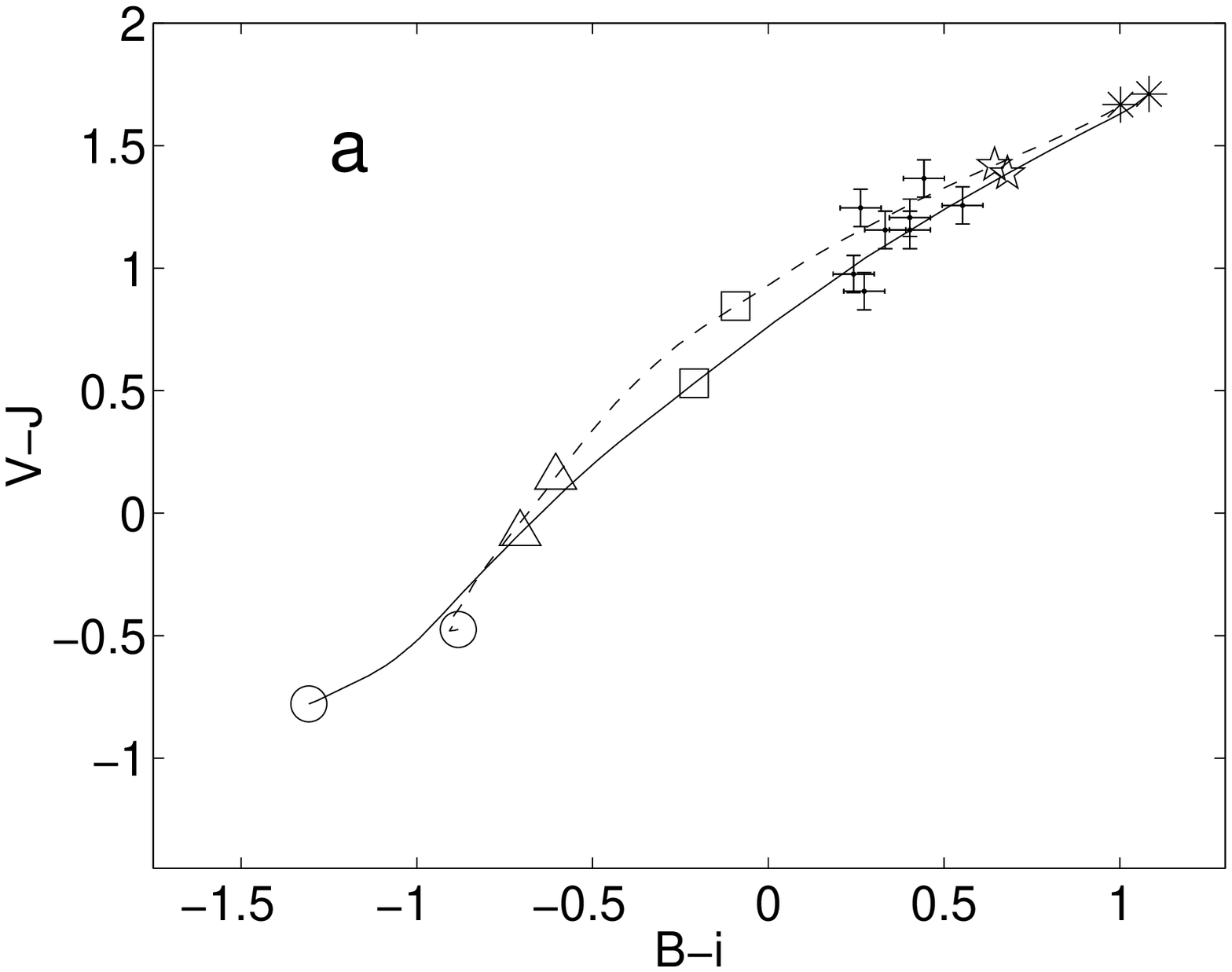} \includegraphics{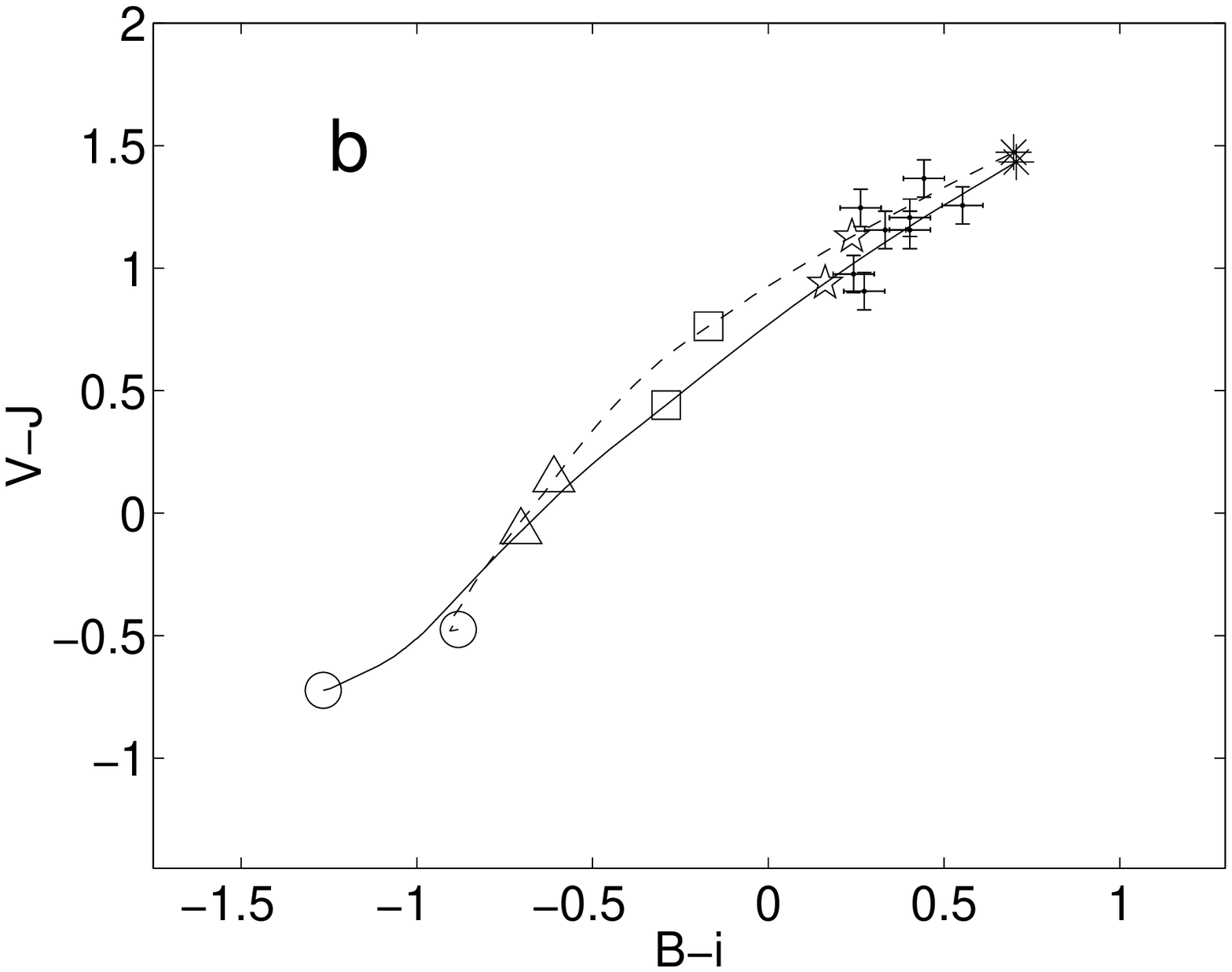}\includegraphics{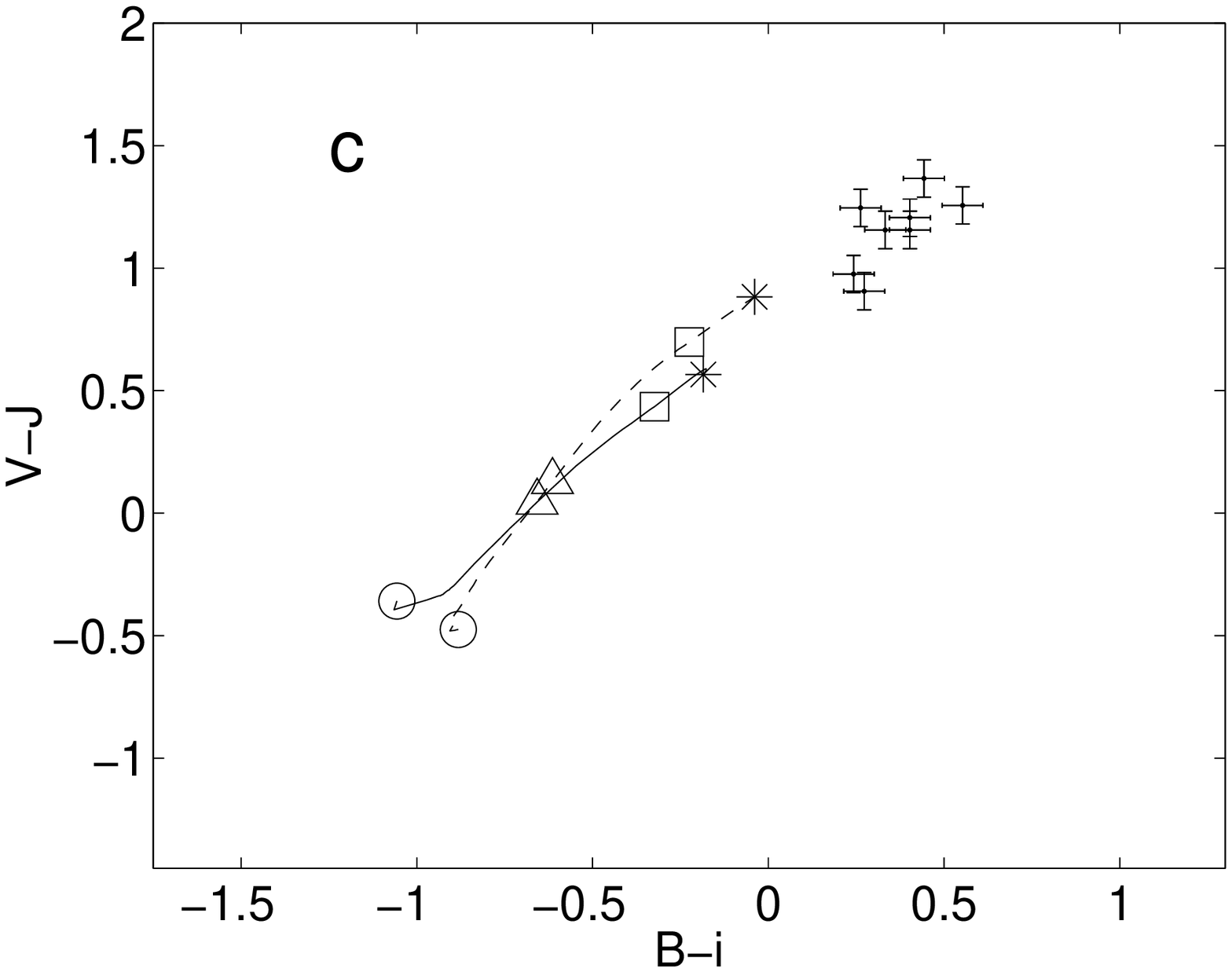}}
\caption[]{The temporal evolution of $B-i$ vs. $V-J$ predicted by Z01 (solid) and P\'EGASE.2 (dashed). Both models assume a metallicity of $Z=0.001$, a Salpeter IMF throughout the mass range 0.08--120 $M_\odot$, a redshift of $z=0$ and an exponential star formation history with {\bf a)} $\tau=1$ Gyr, {\bf b)} $\tau=6$ Gyr and {\bf c)} $\tau=-2$ Gyr. The additional parameter values for Z01 are a filling factor 0.1 and a total gas mass of $10^{10} \ M_\odot$ available for star formation. Markers indicate ages of 1 Myr (circle), 100 Myr (triangle), 1 Gyr (square), 5 Gyr (pentagram, omitted from rightmost plot to avoid cluttering) and 15 Gyr (asterisk) along the model sequences. The colours of the eight objects (corrected for intrinsic extinction using $c=0.15$), for which data exist in all four filters required, are indicated by crosses representing $1\sigma$ error bars.}
\label{BIVJ}
\end{figure*}

\subsection{Model fitting}
For each time step along each of the 540 Z01 evolutionary sequences defined in Table \ref{modelgrid} and the 54 P\'EGASE.2 sequences of Table \ref{modelgrid2}, a least-squares fit is performed for each object between the $N$ observational and model passband fluxes, $f_{\mathrm{obs},i}$ and $f_{\mathrm{mod},i}$, weighted by their photometric uncertainties, $\sigma_i$. An average model-observation residual, $\Delta_\mathrm{mod}$, is then calculated using:
\begin{equation}
\Delta_\mathrm{mod}=\sqrt{\frac{\sum_{i=1}^N \Delta_i^2 w_i}{\sum_{i=1}^N w_i}},
\end{equation}
where $w_i=\sigma_i^{-2}$ and 
\begin{equation}
\Delta_i=\left ( \frac{f_{\mathrm{obs},i}-f_{\mathrm{mod},i}}{f_{\mathrm{obs},i}} \right ).
\end{equation}

To estimate confidence levels on the stellar population parameters derived from this fitting procedure, numerically estimated $1\sigma$ (68\%) and $2\sigma$ (95\%) levels on the model-observation residuals are imposed.  

During the fitting procedure, the observed photometric fluxes in all filters except $R$ are used. The reason for the omission of $R$-band data is the uncertain calibration of low surface brightness objects in the ESO/Uppsala catalogue. When comparing the $B-R$ photometry from the ESO/Uppsala catalogue with the results from CCD photometry of bulge-dominated LSBGs carried out by Beijersbergen et al. (\cite{Beijersbergen et al.}), a rather large scatter in the residuals becomes evident, corresponding to $\sigma_{B-R} \approx 0.3$ mag. For our galaxies which have fainter central regions, the calibration problems may be even more severe.

Although not explicitly used in the fitting procedure, H$\alpha$ data are used to reject all fits with emission line equivalent widths outside the typically observed range of $10 \ \AA \leq \mathrm{EW(H\alpha)}$ $\leq 60 \ \AA$, as discussed in section 3.7. To allow for the uncertainty in the extinction correction, values of $c=0.10$, $c=0.15$ and $c=0.20$ are considered for each galaxy.

\section{Results} 
In Fig.~\ref{BIVJ}a and \ref{BIVJ}b, we present the behaviour predicted by Z01 and P\'EGASE.2 in a diagram of $B-i$ vs. $V-J$ for two of the evolutionary scenarios most successful in reproducing the observed colours of the galaxies under the assumption of a standard Salpeter IMF. The blue LSBGs in our sample are all located inside a small region of this two-colour diagram, indicating that they all share similar properties. Although slight differences between the model predictions of Z01 and P\'EGASE.2 do exist in these filters, both models are able to roughly reproduce the observed colours of our objects. 

The large number of spectral evolutionary sequences considered in our analysis allows several hundreds to thousands of fits of comparable quality. Among these, a wide variety of model parameter values can be found, indicating that the observations considered here can be equally well reproduced by many fundamentally different star formation scenarios. This degeneracy becomes obvious in Fig.~\ref{BIVJ}a and \ref{BIVJ}b, where the colours of our target objects may be reproduced with both star formation histories depicted (exponentially decaying SFRs with $\tau=1$ and 6 Gyr respectively), although at different ages along these evolutionary sequences. In the $\tau=1$ Gyr case (Fig.~\ref{BIVJ}a), all objects appear to have ages somewhere between 1 and 5 Gyr, whereas the $\tau=6$ Gyr case (Fig.~\ref{BIVJ}b) implies that all objects should have ages in excess of  5 Gyr. This degeneracy between age and star formation history is an old problem (see e.g. McGaugh \& de Blok \cite{McGaugh & de Blok}; Gil de Paz \& Madore \cite{Gil de Paz & Madore}), which nonetheless too often tends to be overlooked. The inclusion of near-IR broadband data, which should be a better tracer of old stars, does not appear to improve the situation substantially. A similar situation also occurs for parameters other than the star formation history, and as it turns out, neither the slope nor the upper mass limit of the IMF can be constrained by the available observations.
\begin{figure*}[t]
\resizebox{\hsize}{!}{\includegraphics{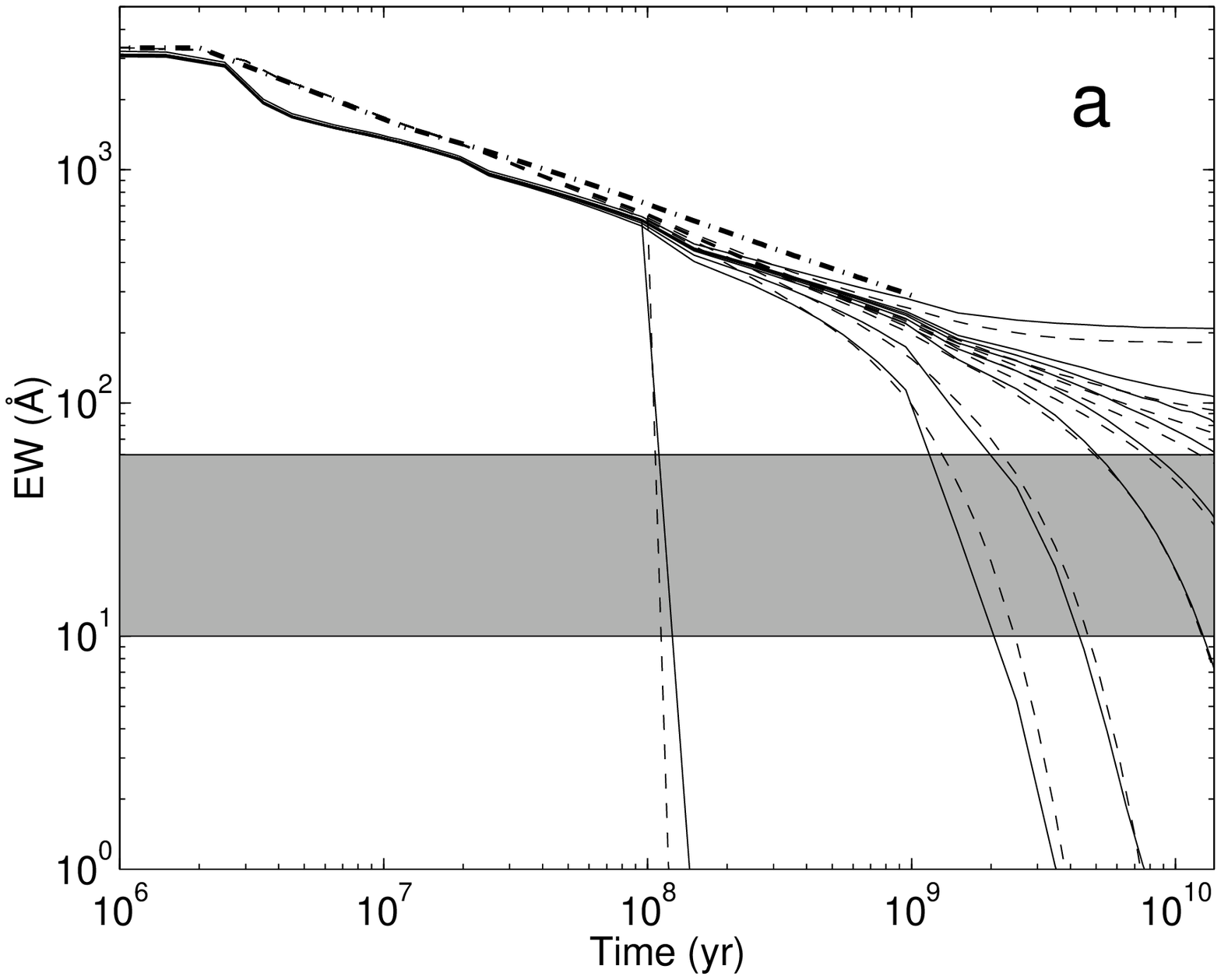}\includegraphics{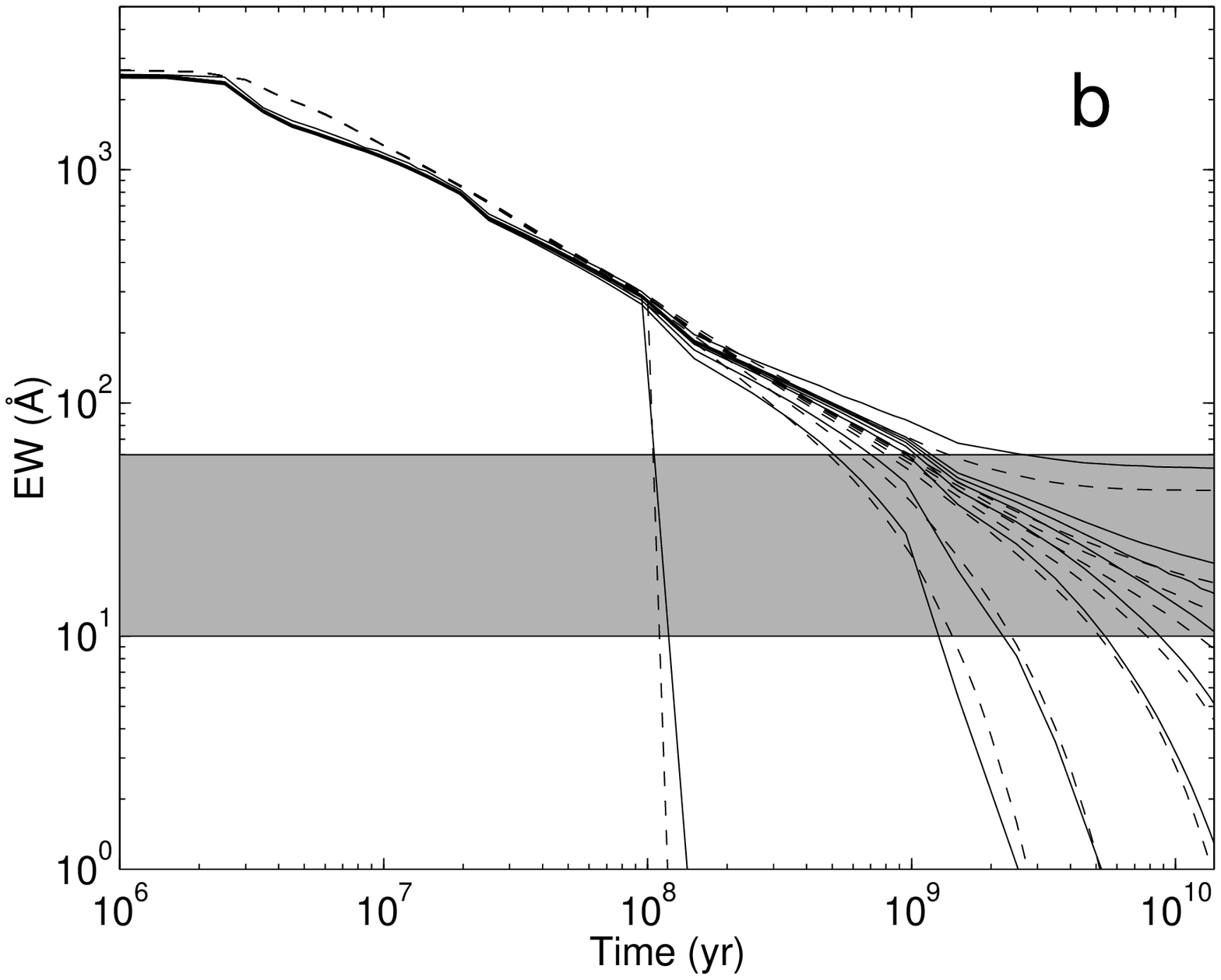}\includegraphics{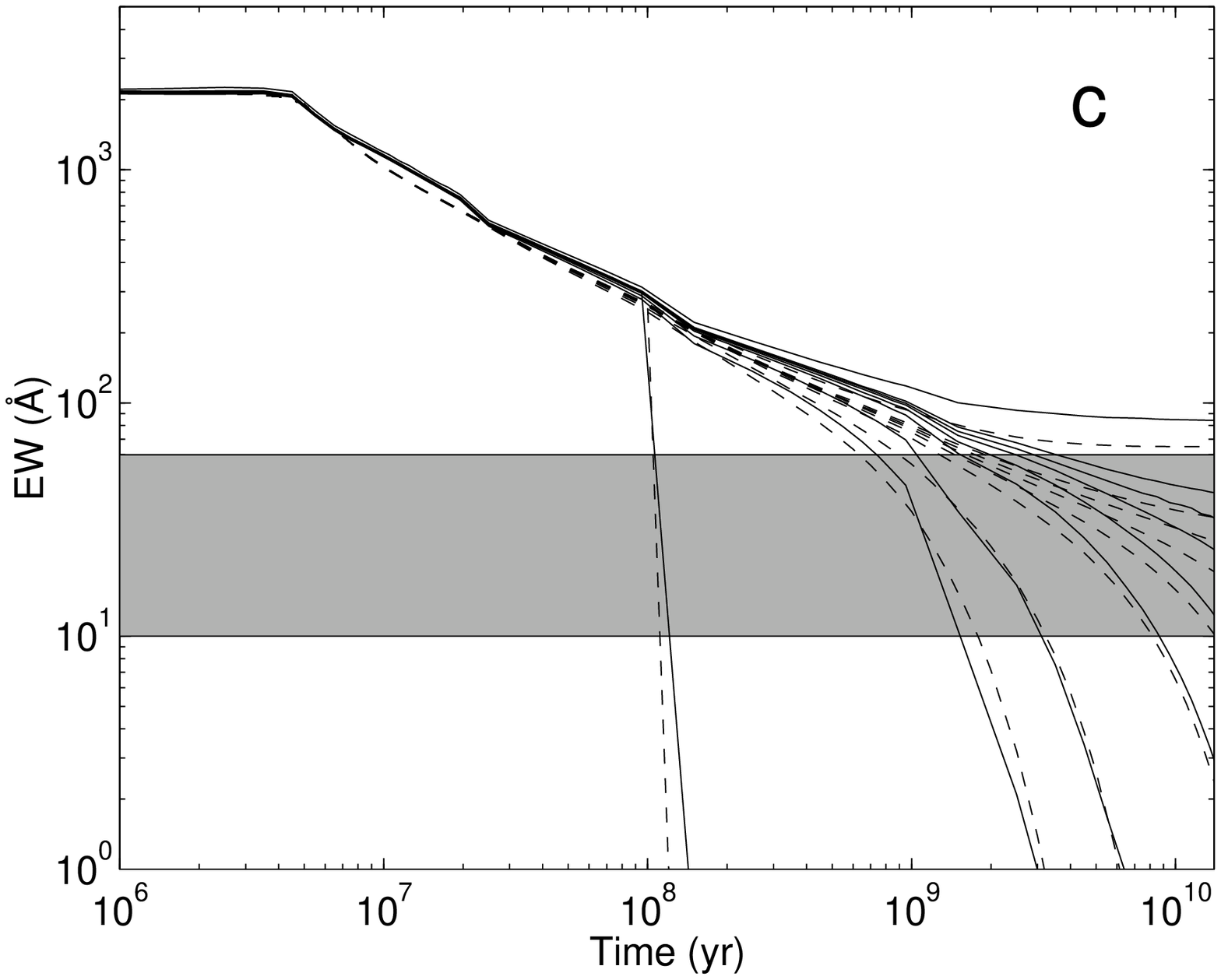}}
\caption[]{The temporal evolution of EW(H$\alpha$) predicted by Z01 (solid lines) and P\'EGASE.2 (dashed) for various SFHs compared to the range observed among the bluest LSBGs (gray area). Both models assume $Z=0.001$. Z01 furthermore assumes a filling factor of 0.1 and $M_\mathrm{tot}=10^{10} \ M_\odot$. The three panels represent different assumptions about the IMF. {\bf a)} A Salpeter IMF with $M_\mathrm{up}=120 \ M_\odot $. {\bf b)} An IMF slope of $\alpha=2.85$ with $M_\mathrm{up}=120 \ M_\odot $. {\bf c)} A Salpeter IMF with an upper mass limit of 40 $M_\odot$. From top to bottom in the right half of each panel, the different lines indicate star formation histories with $\tau=-2$ Gyr, $-15$ Gyr, constant SFR, $\tau=15$ Gyr, 6 Gyr, 3 Gyr, 1 Gyr and 500 Myr. The line dropping rapidly at an age of 100 Myr corresponds to the short burst scenario.  The dash-dotted line running up to 1 Gyr in {\bf a)} represents the predictions of Starburst99 for a constant SFR.}
\label{EW_SFH}
\end{figure*}

\subsection{Star formation history}
Although the available photometry does not appear to impose any strong constraints on the star formation history of the blue LSBGs, their low EW(H$\alpha$) do. In Fig.~\ref{EW_SFH} we compare the temporal evolution of EW(H$\alpha$) predicted by Z01 and P\'EGASE.2 for different SFHs, with the range of EW(H$\alpha$) observed among blue LSBGs.

In the short burst scenario (constant SFR for 100 Myr, followed by no star formation thereafter), much too high values of EW(H$\alpha$) are predicted during the early phase of active star formation. Once star formation has ended, EW(H$\alpha$) on the other hand rapidly drops to values much too low, allowing a time span of only  $\approx 20$ Myr during which this scenario predicts equivalent widths in the observed range. However, even with the wide range of IMFs and metallicities considered in this paper, the bluest LSBGs are not blue enough (in e.g. $B-J$) to be consistent with the colours predicted by the models for this intermediate phase. Hence, the short burst scenario (as defined here, with star formation ending very abruptly) can be ruled out.

In scenarios including constant or increasing star formation rates over cosmological time scales, such as those motivated by predictions for the lowest-mass low surface brightness galaxies in the model of Boissier et al. (\cite{Boissier et al.}), the EW(H$\alpha$) never have time to drop sufficiently to reach the observed range within the age of the universe ($14.1^{+1.0}_{-0.9}$ Gyr; Tegmark et al. \cite{Tegmark et al.}) in the case of an IMF with a Salpeter or flatter (e.g. $\alpha=1.85$) slope. For the most rapidly increasing star formation rate ($\tau=-2$ Gyr) the same IMFs furthermore produce colours which are much too blue for our objects, as demonstrated for a Salpeter IMF in Fig.~\ref{BIVJ}c. Both these conclusions are confirmed by P\'EGASE.2, indicating that, under the assumption of a standard IMF, these star formation scenarios cannot be valid for the low-metallicity, low surface brightness galaxies considered here. The idea of a highly variable past SFH, which was advanced by Boissier et al. (\cite{Boissier et al.}) to explain the presence of red LSBGs, does not appear attractive for our objects, as discussed in Section 3.8. In Fig.~\ref{EW_SFH}a, the EW(H$\alpha$) evolution during the first Gyr predicted by Starburst 99 (Leitherer et al. \cite{Leitherer et al.}; version 4.0), for a constant SFR scenario, is included to demonstrate that the problem is not due to unrealistically high Lyman continuum fluxes in Z01 and P\'EGASE.2, which employ stellar atmosphere models for high-mass stars considerably less sophisticated than those of Starburst 99. If anything, Z01 and P\'EGASE.2 appear to {\it underpredict} the equivalent widths. 

Although the EW(H$\alpha$) predicted by Z01 and P\'EGASE.2 neglect the effect of underlying absorption by the stellar population, this may have increased the predictions by no more than $\approx5$ \AA{} (R\"onnback \& Bergvall \cite{ronnback2}) compared to the observations for these objects. Selective extinction due to emission lines originating in regions more dust-rich than those responsible for the stellar continuum (Calzetti et al. \cite{Calzetti et al.}) may furthermore lead to corrections of only $\lesssim 25\%$ to the observed EW(H$\alpha$) for the low levels of extinction (R\"onnback \cite{Rönnback}) observed in our targets. 

The very high EW(H$\alpha$) predicted by scenarios including constant or increasing star formation rates may nonetheless be remedied if the slope of the IMF is significantly steeper (e.g. $\alpha=2.85$), or the upper mass limit substantially lower (e.g. $M_\mathrm{up}=40 \ M_\odot$) than typically assumed, as shown in Fig.~\ref{EW_SFH}b and \ref{EW_SFH}c, respectively. Lowering $M_\mathrm{up}$ does not necessarily contradict the observed (N/O) and (O/H) abundances of blue LSBGs (R\"onnback \cite{Rönnback}), since this alteration has only a modest impact on the abundances of these elements (Olofsson \cite{Olofsson}). A bottom-heavy (high $\alpha$) IMF could help to explain the high $M/L$ of LSBG disks claimed by Fuchs (\cite{Fuchs}), as demonstrated by Lee et al. (\cite{Lee et al.}), but may produce too low (O/H) abundances. A high $\alpha$ just for high-mass ($>1 M_\odot$) stars would on the other hand be consistent with recent models for the field-star IMF (Kroupa \& Weidner \cite{Kroupa & Weidner}).  A more detailed investigation of what constraints the EW(H$\alpha$) and chemical abundances can impose on the SFHs and IMFs of LSBGs (and not just the bluest objects) would be very valuable.

An alternative way to reconcile the median observed EW(H$\alpha$) with constant or increasing SFRs could be to reduce the hydrogen-ionizing flux shortward of the Lyman limit by advocating dust extinction. In order for this to work, the required Lyman continuum depletion factor would however have to be higher than 2.5. It is not obvious that the amount and spatial distribution of dust required to achieve this is at all realistic for blue LSBGs, whose star-forming regions show only modest signs of reddening. 

\begin{table*}[t]
\caption[]{The range in age, $M/L_V$ and $M_\star$ allowed by the acceptable fits to the observations at the $1\sigma$ and $2\sigma$ levels using the Z01 grid of evolutionary sequences defined in Table \ref{modelgrid} and the P\'EGASE.2 grid of evolutionary sequences defined in Table \ref{modelgrid2}. Only the results from sequences assuming a Salpeter IMF with $M_\mathrm{up}=120 \ M_\odot$ have been included. Missing entries indicate the lack of acceptable fits at a given confidence level or, in the case of $M_\star$, the lack of redshift data for a particular object.  To obtain values of $M/L_V$ and $M_\star$ consistent with the observed IMF of Kroupa (\cite{Kroupa}), these values should be multiplied by a factor of $\approx 0.6$, as discussed in section 4.3.} 
\begin{flushleft}
\begin{tabular}{l|llll|llll|lllll} 
\hline
	& Age  & & & & $M/L_{V}$ & & & & $\log$ & & & \\
	& (Gyr)          & & & &           & & & & $M_\star \ (M_\odot)$ & & & \\
	& Z01	    &           & P\'EG2&  & Z01 & &  P\'EG2 & & Z01 & &  P\'EG2 &\\
Object	& $1\sigma$ & $2\sigma$ & $1\sigma$ & $2\sigma$ & $1\sigma$ & $2\sigma$  & $1\sigma$ & $2\sigma$ & $1\sigma$ & $2\sigma$ & $1\sigma$ & $2\sigma$ &\\
\hline
\hline
\object{ESO 074-16} & 2--15 & 1--15 & 1--15 & 1--15 & 0.8--2.4 & 0.8--2.8 & 0.6--1.9 & 0.5--1.9 & 8.9--9.4 & 8.9--9.5 & 8.8--9.3 & 8.7--9.3 \\
\object{ESO 084-41} & 2--15 & 1--15 & 2--8 & 1--15 & 0.8--2.2 & 0.8--2.6 & 0.6--1.3 & 0.5--1.9 & -        & -        & -       & -       \\
\object{ESO 146-14} & -     & - & 1--2 & 1--5  & -        & - & 0.6 & 0.6--1.0 & -        & - & 8.7     & 8.6--8.9\\
\object{ESO 288-48} & 2--9 & 1--9 & 1--8 & 1--15 & 0.8--1.8 & 0.8--2.2 & 0.5--1.3 & 0.5--1.8 & 7.8--8.2 & 7.8--8.2 & 7.6--8.0& 7.6--8.2\\
\object{ESO 462-36} & -     & 2--15 & 2--15 & 1--15 & -        & 1.1--2.4 & 0.6--2.0 & 0.6--2.4 & -        & -        & -       & -       \\
\object{ESO 505-04} & 2     & 2--10 & 1--13 & 1--15 & 0.8	  & 0.8--2.0 & 0.6--1.6 & 0.5--1.9 & 8.8--8.9 & 8.8--9.2 & 8.7-9.1 & 8.7--9.2\\
\object{ESO 546-09} & 2--15 & 2--15 & -     & 2--15$^\mathrm{a}$  & 0.9--3.1 & 0.9--3.5 & - & 0.6--2.2$^\mathrm{a}$  & 9.4--9.8 & 9.3--9.9 & -       & 9.2--9.7$^\mathrm{a}$ \\
\object{ESO 546-34} & 2     & 2     & -     & 1--2$^\mathrm{a}$ & 0.8--1.0 & 0.8--1.1 & -        & 0.5--0.6$^\mathrm{a}$ & 8.3--8.4      & 8.3--8.5 & -       & 8.2$^\mathrm{a}$ \\
\object{ESO 602-26} & -     & 2--3  & -     & 1--2$^\mathrm{a}$ & - & 0.8--1.2 & -        & 0.5--0.7$^\mathrm{a}$  & -        & -        & -       & -       \\
\hline
\end{tabular}
a) $3\sigma$ confidence levels used, due to lack of acceptable fits within the $2\sigma$ limit
\label{constraints_Z01+Peg2}
\end{flushleft}
\end{table*}
\begin{table*}[t]
\caption[]{The allowed range of stellar $M/L$ ratios among the blue LSBGs in our sample for which the Z01 and P\'EGASE.2 models are able to provide adequate fits at levels of $1\sigma$  and $2\sigma$, respectively. Only the results  from model sequences assuming a Salpeter IMF with $M_\mathrm{up}=120 \ M_\odot$ have been used. To obtain values of $M/L$ consistent with the IMF of Kroupa (\cite{Kroupa}), these values should be multiplied by a factor of $\approx 0.6$, as discussed in section 4.3.} 
\begin{flushleft}
\begin{tabular}{llllllllll} 
\hline
Model	& Confidence     & $M/L_{B}$ & $M/L_{V}$ & $M/L_{R}$ & $M/L_{i}$ & $M/L_{I}$ & $M/L_{J}$ & $M/L_{H}$ & $M/L_{K}$\\
\hline
\hline
Z01		& $1\sigma$ & 0.6--2.5 & 0.8--3.1 & 0.9--2.9 & 0.9--2.7 & 0.9--2.9 & 0.8--2.4 & 0.7--2.2 & 0.7--2.0\\
		& $2\sigma$ & 0.6--3.0 & 0.8--3.5 & 0.9--3.3 & 0.9--3.1 & 0.9--3.2 & 0.8--2.6 & 0.7--2.4 & 0.7--2.2\\
P\'EGASE.2 	& $1\sigma$ & 0.4--1.7 & 0.5--2.0 & 0.6--1.9 & 0.6--2.0 & 0.7--2.1 & 0.5--1.5 & 0.4--1.4 & 0.4--1.3\\
		& $2\sigma$ & 0.4--2.1 & 0.5--2.4 & 0.6--2.3 & 0.6--2.3 & 0.7--2.4 & 0.5--1.7 & 0.4--1.6 & 0.4--1.4\\
\hline
\end{tabular}
\label{constraints_ML}
\end{flushleft}
\end{table*}
\subsection{Ages}
Here, the absolute age (or age, for short) of a galaxy is defined as the time passed from the first onset of significant star formation processes, i.e. the start of the assumed SFH. The average age of a galaxy, which will be discussed in relation to previous investigations on LSBGs, is instead defined as the average age of all stars ever formed in the system. Since stellar remnants are also included in this inventory, the average age is a function of SFH and absolute age only.
 
The range of absolute ages among the acceptable fits generated by Z01 and P\'EGASE.2 for our blue LSBGs are presented in Table~\ref{constraints_Z01+Peg2}. These intervals all assume a Salpeter IMF with $M_\mathrm{up}=120 \ M_\odot$.  

Both models agree that the observations for most of the objects in our sample can be equally well reproduced by ages anywhere in the range from 1 or 2 Gyr, up to 15 Gyr, which is the highest age considered here. The ages are therefore not well constrained by the available observations, despite the use of near-IR data. 

What about the average ages? In Bell et al. (\cite{bell}), colour-colour diagrams are presented with model grids containing variations in metallicity and star formation history. For an assumed absolute age of the galaxies (12 Gyr, in their case), they show that the average stellar age can in fact be constrained to impressive levels using broadband photometry in four filters only. As we demonstrate in Fig.~\ref{Meanagetau}, problems do however arise when the absolute age is no longer assumed to be known. In this case, the grid shifts as a function of absolute age, so that e.g. the colours of the group of blue LSBGs around $B-V\approx 0.4$ mag, $V-J\approx 1.2$ mag in Fig.~\ref{Meanagetau} may be equally well represented by an absolute age of 3.0 Gyr, an average age of 2.2 Gyr and a star formation history with $\tau=1.0$ Gyr (left panel) as by an absolute age of 13.0 Gyr, an average age of 7.4 Gyr and $\tau=15.0$ Gyr (right panel). Hence, neither the absolute nor average ages are particularly well constrained. For the LSBGs studied by Bell et al. (\cite{bell}), the uncertainty is likely to be even more severe due to the unknown metallicities of their objects. 
\begin{figure*}[t]
\resizebox{\hsize}{!}{\includegraphics{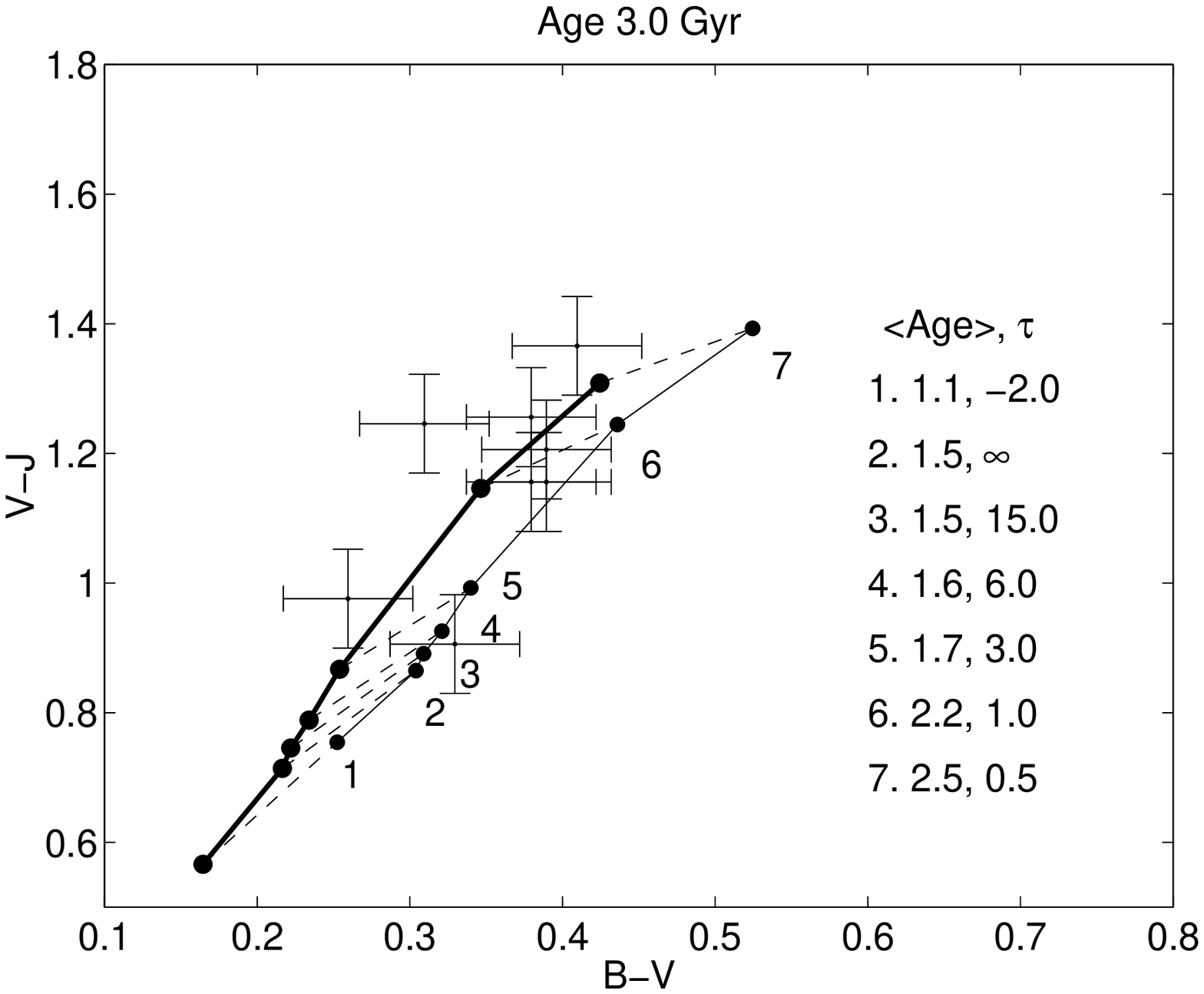}\includegraphics{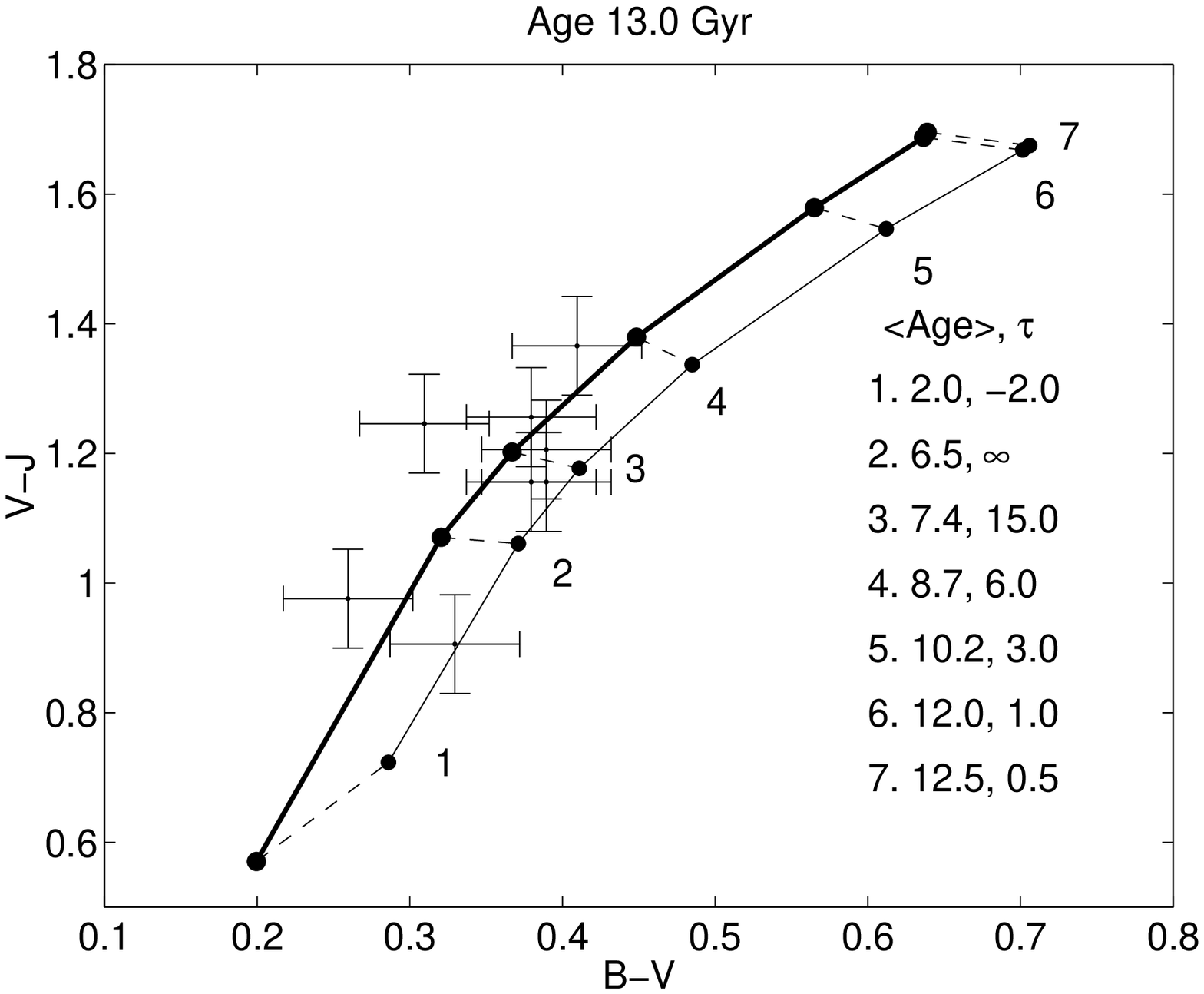}}
\caption[]{Diagrams of $B-V$ vs. $V-J$, showing the predicted positions of stellar populations at absolute ages of 3 Gyr (left panel) and 13 Gyr (right panel). In each panel, lines corresponding to metallicities of $Z=0.001$ (thick solid) and $Z=0.004$ (thin solid) connect the colours predicted for different star formation histories, parametrized by $\tau$ in $\mathrm{SFR}(t)\propto \exp{(-t/\tau)}$. Each combination of absolute age and $\tau$ corresponds to the listed average stellar population ages, $<$Age$>$. Dashed lines connect predicted populations with the same SFH but different $Z$. The colours (corrected for intrinsic extinction using $c=0.15$) of the eight objects for which data exists in all three filters are indicated by crosses representing $1\sigma$ error bars. A Salpeter IMF with $M_\mathrm{up}=120 \ M_\odot$ is assumed.}
\label{Meanagetau}
\end{figure*}

From the number of missing entries in Table~\ref{constraints_Z01+Peg2}, it is evident that both Z01 and P\'EGASE.2 have some trouble explaining the observed properties of certain galaxies even at the $2\sigma$ level, although the problematic objects are different for the two models. Both Z01 and P\'EGASE.2 identify about three objects each which are better fitted by young ages ($\leq 3$ Gyr), but these are also the most difficult galaxies to fit simultaneously with both codes. Therefore, the two models cannot be said to convincingly confirm each other's detections of young galaxies. 

\subsection{Stellar mass-to-light ratios and population masses}
The mass-to-light ratios ($M/L_F$) used here are defined as $\frac{M}{M_\odot}/\frac{L_F}{L_{\odot,F}}$, where $F$ represents a particular filter and $M$ the mass locked up in stars or stellar remnants. From the $V$-band mass-to-light ratios ($M/L_V$,) and the $V$-band luminosities ($L_V$) predicted by the model fits, stellar population masses ($M_\star$) are computed. 
Table~\ref{constraints_Z01+Peg2} lists the range of $M/L_V$ and $M_\star$ allowed by the acceptable fits generated by Z01 and P\'EGASE.2 for our target objects. At the $2\sigma$ level, $M/L_V=0.5$--3.5 and $M_\star\approx 10^7$--$10^{10} \ M_\odot$. These constraints all assume a Salpeter IMF with $M_\mathrm{up}=120 \ M_\odot$.  

The derived $M/L$ ratios are of course subject to substantial uncertainties due to the poorly constrained scaling of the IMF. A single-valued Salpeter power-law is generally believed to produce too many low-mass stars (e.g. Kroupa \cite{Kroupa}, Chabrier \cite{Chabrier}), and scaling relations of the type $(M/L)\approx X\ (M/L)_\mathrm{Salpeter}$ have been suggested. Both direct integration of the observed mass function (Kroupa \cite{Kroupa}) and dynamical arguments (McGaugh \cite{McGaugh3}) favour a value of $X \approx 0.6$ (factor relating a Salpeter IMF with a lower mass limit of $M_\mathrm{low}=0.08 \ M_\odot$, as adopted here, to the Kroupa \cite{Kroupa} IMF), and it may therefore be suitable to multiply the $M/L$ ratios of Table~\ref{constraints_ML} by a factor of this order.

In many contexts, e.g. in rotation-curve decompositions and in the study of the Tully-Fisher relation, it is advantageous to use the filter passband for which the stellar $M/L$ ratios are the least vulnerable to uncertainties in age, dust, metallicity and star formation history. In de Jong (\cite{de Jong}) and Bell \& de Jong (\cite{Bell & de Jong}) it was argued that the $K$-band should be superior in this respect. In Table~\ref{constraints_ML} we present the range of $M/L$ in filters $BVRiIJHK$ allowed by Z01 and P\'EGASE.2 for our target objects. Due to differences in the stellar evolutionary tracks adopted by the two codes (discrepancies between Geneva and Padova tracks, compounded by the non-standard treatment of horizontal branch morphologies used in Z01), Z01 is somewhat underluminous at ages above 1 Gyr and systematically predicts higher $M/L$ ratios. 

Although we do find that the allowed range of $M/L$ typically becomes smaller with increasing central filter wavelength for each spectral evolutionary model separately, this advantage is counterbalanced by the systematic differences in the $M/L$-predictions between Z01 and P\'EGASE.2. The ratio of highest to lowest allowed $M/L$ among both models instead has a minimum around the $i$ or $I$-band. This indicates that while stellar populations theoretically are expected to display a narrow range of near-IR $M/L$ ratios, this effect is partly cancelled by model uncertainties associated with the treatment of late stages of stellar evolution, which have a significant impact on the spectral energy distribution in the near-IR for old stellar populations (see e.g. Portinari \cite{Portinari}). A similar conclusion about the superiority of the $I$ band in  stellar-population $M/L$ studies has previously been reached by Worthey (\cite{Worthey}), by considering the age and metallicity sensitivity of $M/L$ in different passbands.

\subsection{The effects of a peculiar IMF}
\begin{figure*}[t]
\resizebox{\hsize}{!}{\includegraphics{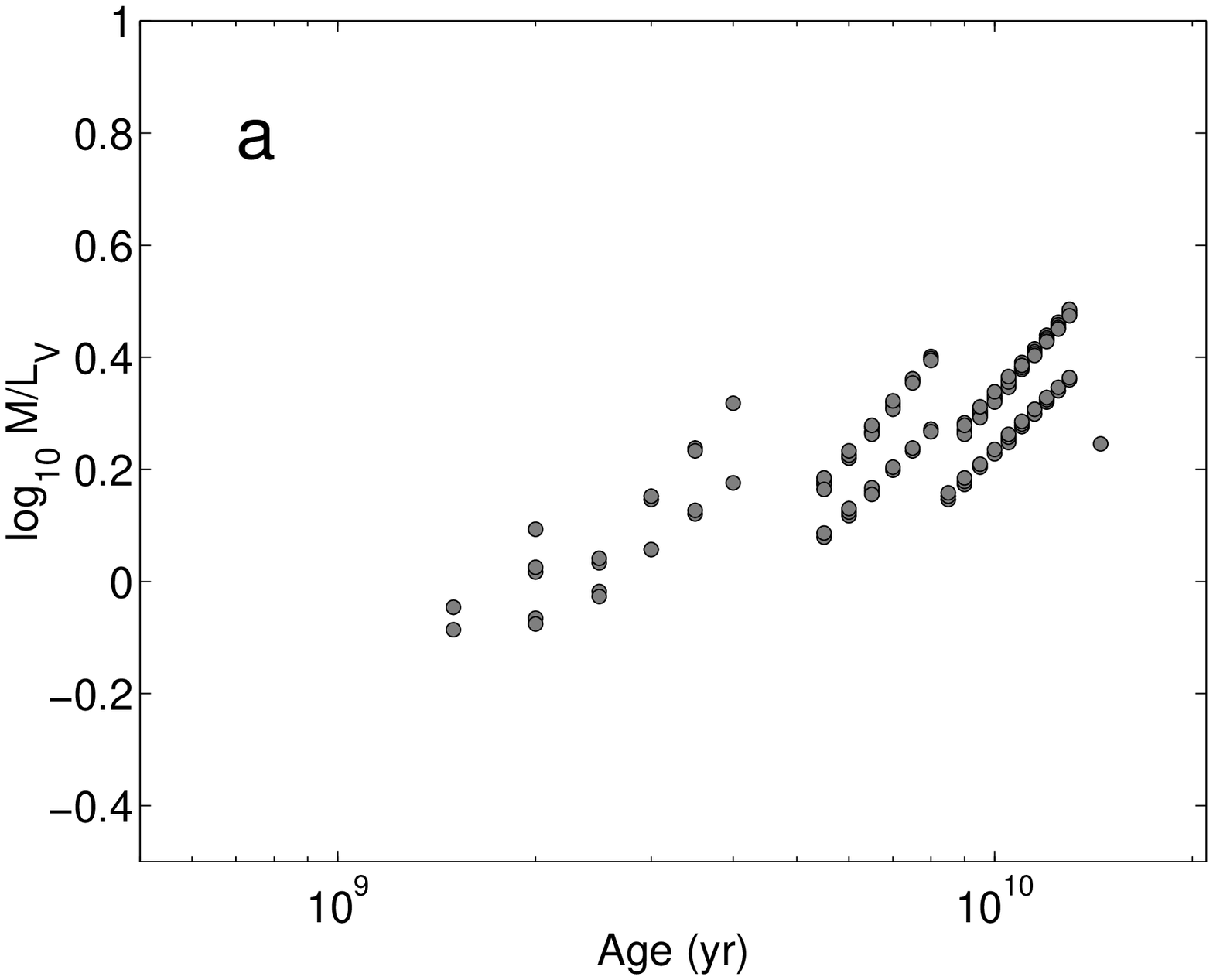}\includegraphics{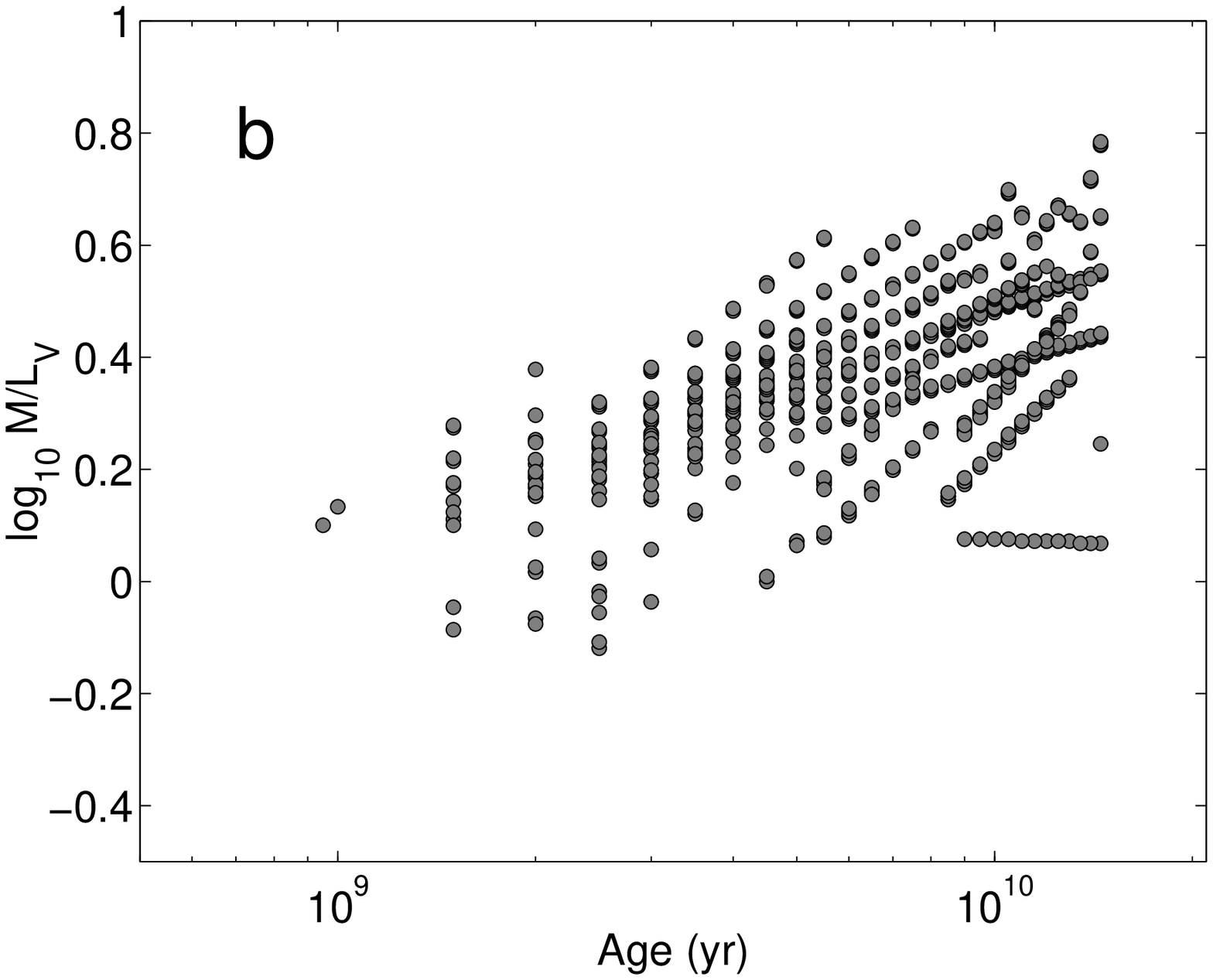}\includegraphics{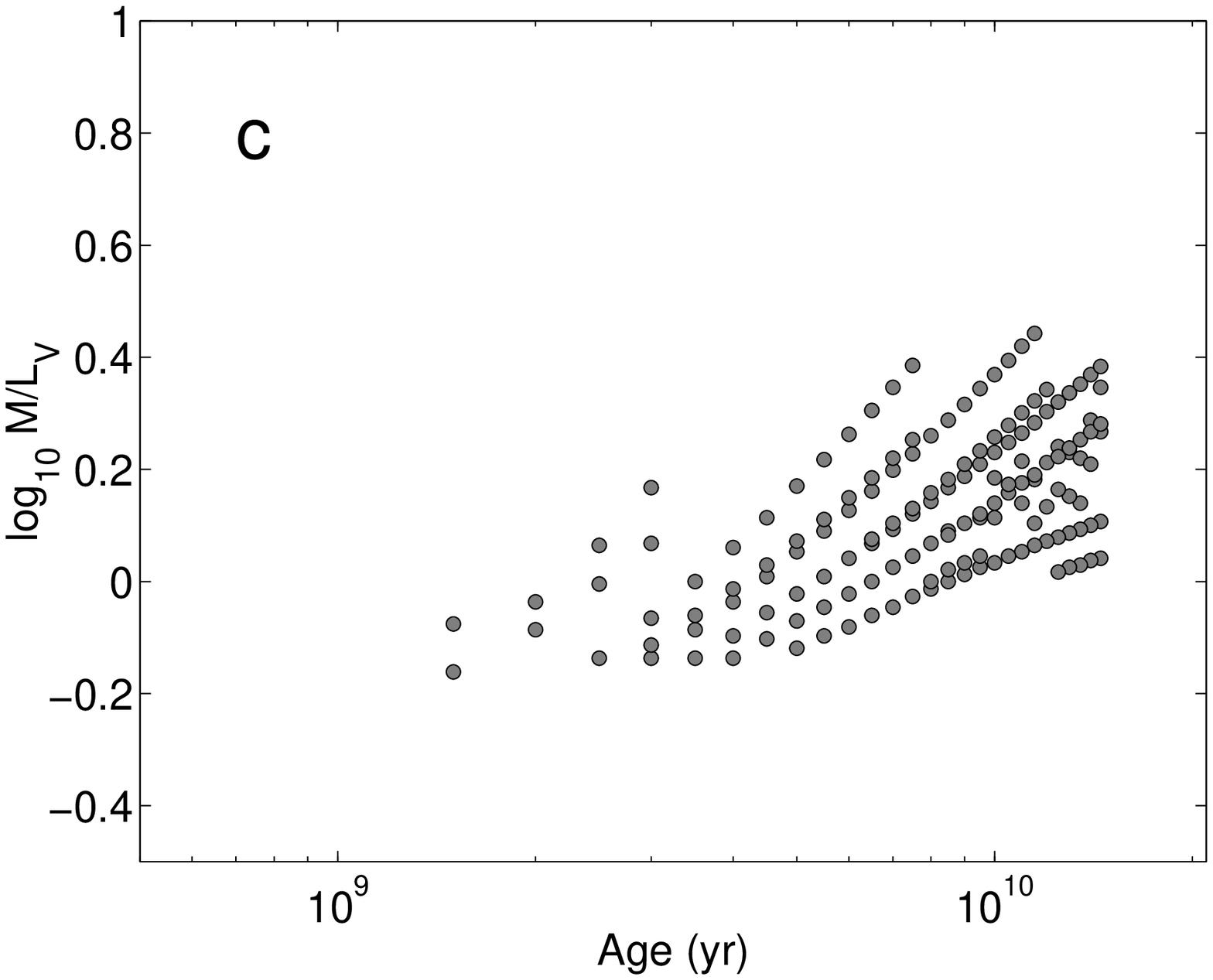}}
\resizebox{\hsize}{!}{\includegraphics{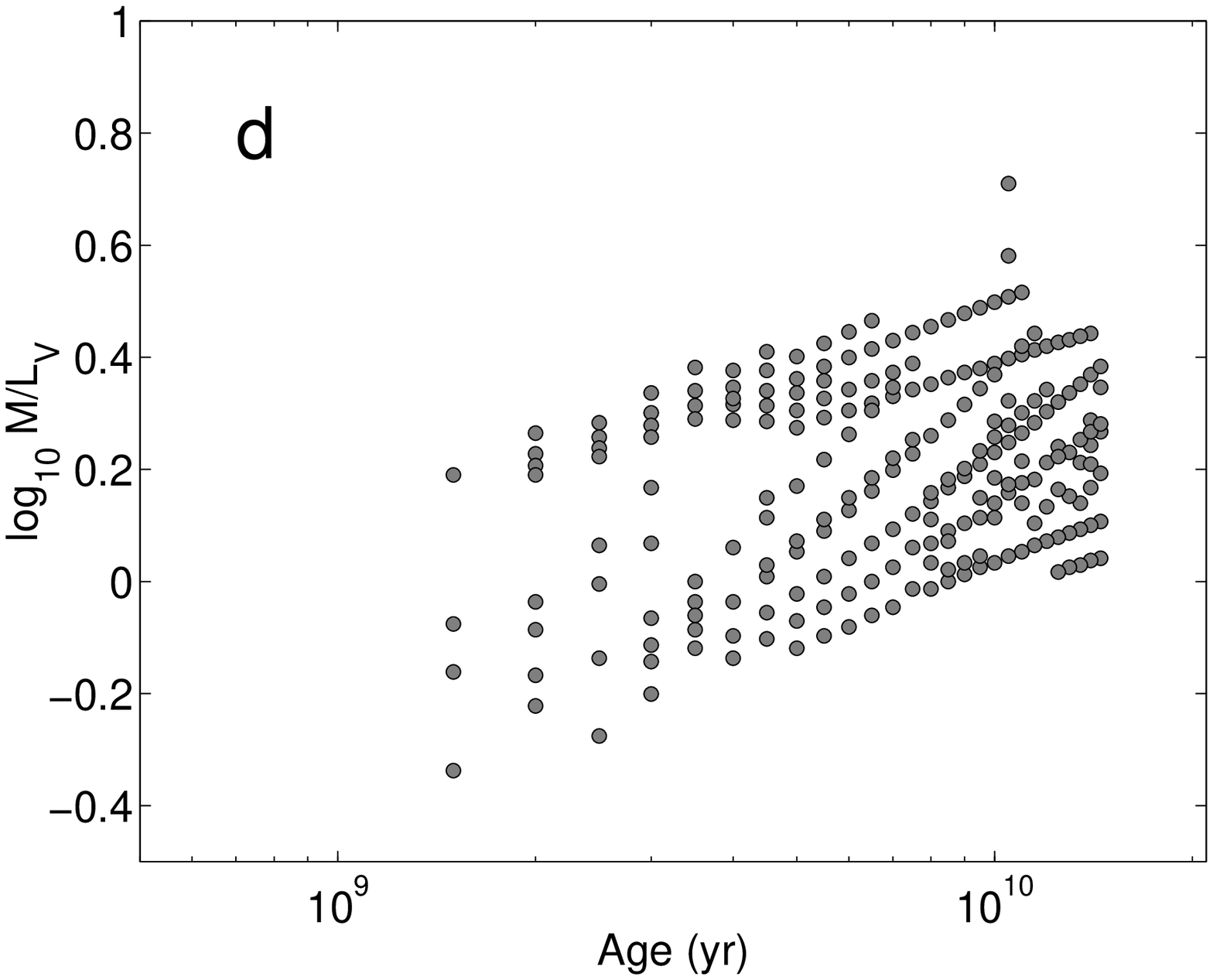}\includegraphics{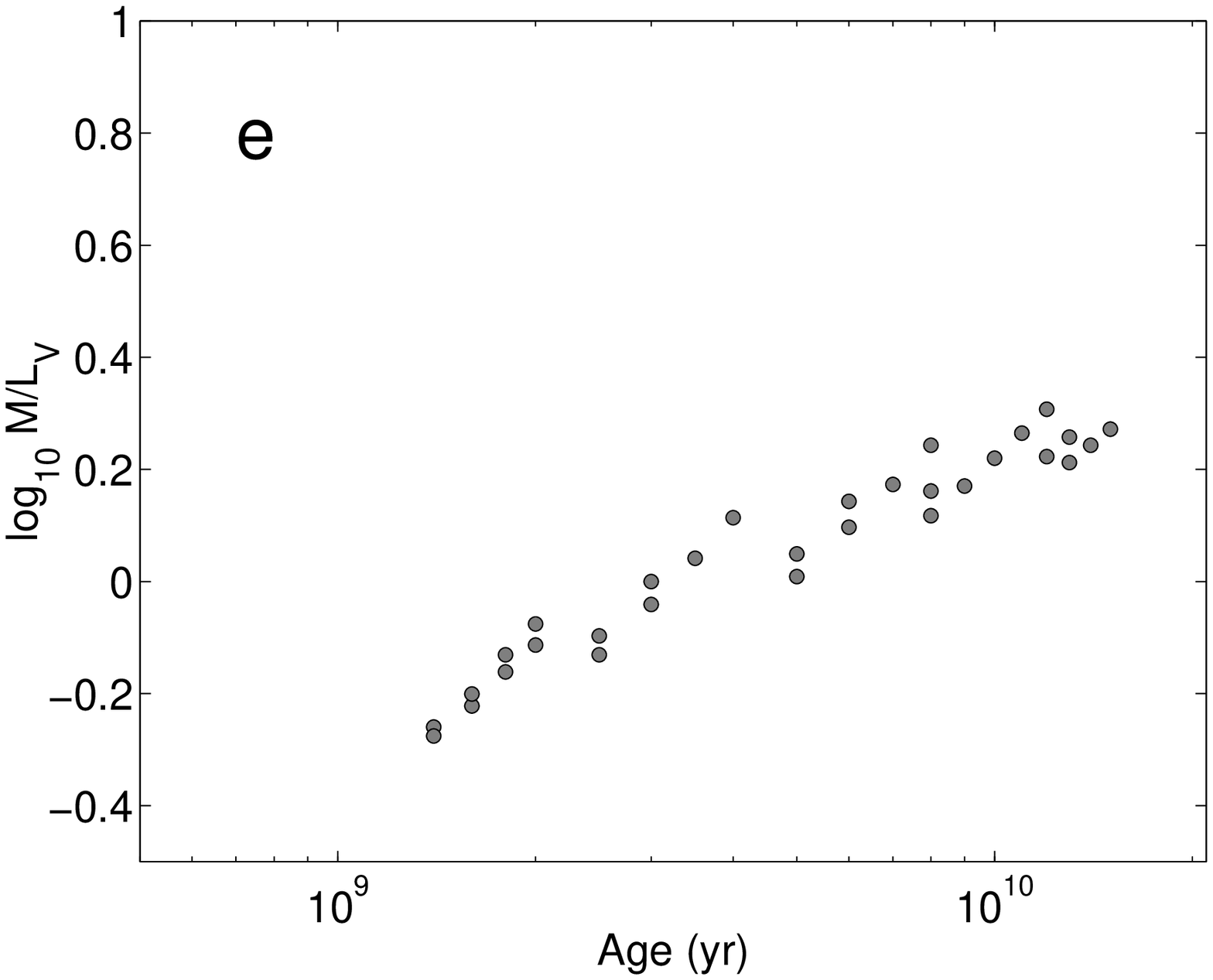}\includegraphics{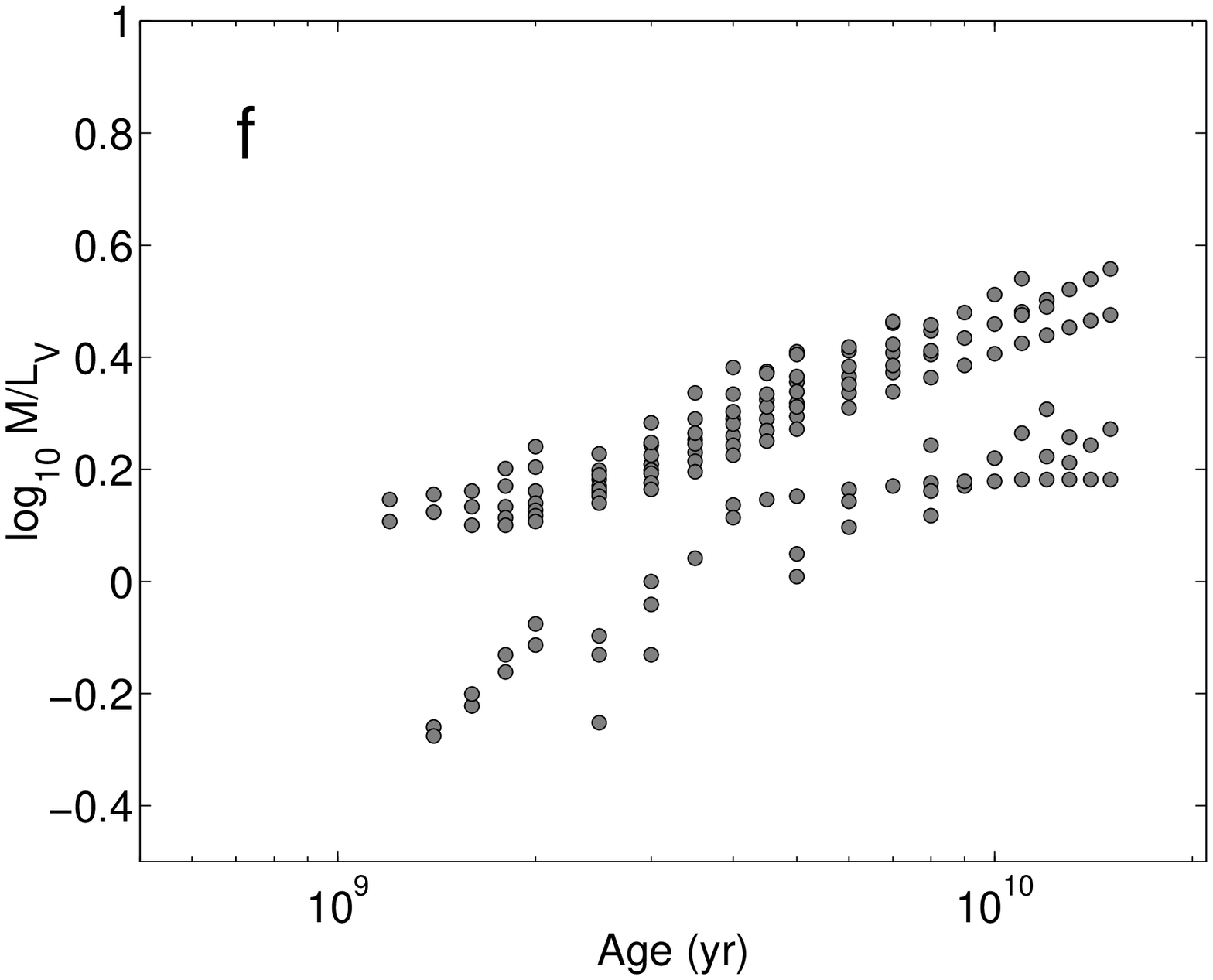}}
\caption[]{The allowed ages and $M/L_V$ of acceptable fits (markers) for both Z01 and P\'EGASE.2 at the 1$\sigma$ level, under various assumption about the IMF: {\bf a)} $M_\mathrm{up}=120 \ M_\odot$, $\alpha=2.35$ for Z01, {\bf b)} $M_\mathrm{up}=120 \ M_\odot$, $\alpha=1.85$--2.85 for Z01, {\bf c)} $M_\mathrm{up}=40 \ M_\odot$, $\alpha=2.35$ for Z01, {\bf d)} $M_\mathrm{up}=40 \ M_\odot$, $\alpha=1.85$--2.85 for Z01, {\bf e)} $M_\mathrm{up}=120 \ M_\odot$, $\alpha=2.35$ for P\'EGASE.2, {\bf f)} $M_\mathrm{up}=120 \ M_\odot$, $\alpha=1.85$--2.85 for P\'EGASE.2. }
\label{MLage}
\end{figure*}
What happens to the derived quantities if the assumption of a Salpeter IMF with $M_\mathrm{up}=120 \ M_\odot$ is relaxed? In Fig.~\ref{MLage}, all the combinations of age and $M/L_V$ allowed for the blue LSBGs at the $1\sigma$ level by Z01 as well as P\'EGASE.2 are plotted, under various assumptions about the IMF.

When keeping $M_\mathrm{up}=120 M_\odot$, but allowing the slope of the IMF to deviate from the Salpeter value, the derived span of $M/L_V$ is increased upwards, allowing $M/L_V$ as high as 6.1 (Z01) and 3.6 (P\'EGASE.2) at the $1\sigma$ level. The change is almost entirely due to the inclusion of bottom-heavy ($\alpha=2.85$) IMFs. In this scenario, high $M/L_V$ arise naturally because of the increased number of low-mass, low-luminosity stars. Since SFHs with constant or increasing SFRs produce acceptable EW(H$\alpha$) in this case (see Fig.~\ref{EW_SFH}), low $M/L_V$ at high ages are however also allowed. 

Assuming an upper mass limit below 120 $M_\odot$ alters the shape of the allowed region in the plot of age versus $M/L_V$, but has a modest impact on the extremum values. Analogous to the case with $\alpha=2.85$, low $M/L_V$ are produced also at high ages. Since the number of low-mass stars is almost unaffected in this case, no high $M/L_V$ are however predicted.

The most notable effect of simultaneously setting $M_\mathrm{up}=40 \ M_\odot$ and allowing $\alpha=1.85$--2.85 is that $M/L_V$ as low as $M/L_V=0.44$ are generated. 

\subsection{Comparison to other investigations}
Despite the use of both optical and near-IR broadband photometry, the age constraints derived here (2--15 Gyr) are essentially equivalent to those derived mainly from optical photometry by R\"onnback (\cite{Rönnback}) and Bergvall \& R\"onnback (\cite{bergvall0}), indicating that the addition of near-IR photometry is inadequate to impose useful constraints on the ages without prior knowledge of the star formation history. The mass estimates for the stellar component presented here are also of the same order of magnitude as those derived in Bergvall \& R\"onnback (\cite{bergvall0}) and R\"onnback (\cite{Rönnback}), although estimates vary significantly for individual objects due to some of the very extreme IMFs allowed in these older papers.

The majority of LSBGs adhere to the Tully-Fisher relation, indicating that their total $M/L$ ratios are higher than in normal disk galaxies and that these objects are more dark matter dominated (e.g. McGaugh \& de Blok \cite{McGaugh & de Blok 2}). This feature has made LSBGs popular targets for measuring the dark halo density profile. In the procedure of rotation curve decomposition common when trying to distinguish between different density profiles, some $M/L$ ratio believed to be representative of the stellar component is typically assumed. Often, this ratio is adopted from colour-$M/L$ relations (e.g. Bell \& de Jong \cite{Bell & de Jong}), which are based on spectral evolutionary scenarios. The $M/L$ ratios derived this way do however depend on the assumptions regarding the properties of the stellar populations of LSBGs, such as the age, metallicity and star formation history used when constructing the relation. 

When the colour-$M/L$ relations from  Bell \& de Jong (\cite{Bell & de Jong}) are applied to our targets ($B-V=0.24$--0.46 mag after correction for intrinsic extinction $c=0.1$--0.2), stellar $M/L$ of $M/L_V= 0.5$--1.6 and $M/L_K=0.5$--0.8 (from their Table 3, after correcting by a factor of 1.57 for differences in the scaling of the Salpeter IMF) are predicted,  which is consistent with the best-fit $M/L_V=0.5$--3.1 and $M/L_K=0.4$--2.0 derived here. Our range of allowed $M/L$ is more generous, but this is expected, since these Bell \& de Jong predictions only take into account the scatter among {\it average} colour-$M/L$ relations derived from different evolutionary scenarios, whereas our predictions include the {\it total} scatter due to the fitting procedure as well as systematic differences between two spectral evolutionary models. There is furthermore no indication that the $M/L_K$ values derived by us are in conflict with the maximum disk values (Fig. 6 in Bell \& de Jong \cite{Bell & de Jong}) inferred from observations of LSBGs at the bluest colours.

\section{Discussion}
Although the available data appear insufficient to claim a young age for the blue LSBG population, ages of 2--5 Gyr are certainly allowed for all the galaxies in our sample and possibly favoured for a few objects. In the framework of an $\Omega_\mathrm{M}=0.3$, $\Omega_\Lambda=0.7$ ($\Lambda$CDM) universe, an age this low would correspond to a formation redshift of $z\leq 0.5$. According to the hydrodynamic galaxy formation models of Nagamine et al. (\cite{Nagamine et al.}), only a very small fraction of galaxies with present mass in stars lower than $10^{10} \ M_\odot$ (such as the blue LSBGs in our study) should form at such low redshifts in this cosmology. 

To estimate the severity of this potential problem, the fraction of blue LSBGs in the local universe should be compared to the fraction of objects with similar masses formed at low redshift in the hierarchical models.

From the ESO-Uppsala catalogue (Lauberts \& Valentijn \cite{lauberts}), one may estimate that the blue LSBGs obtained with our selection criteria constitute $\approx 18\%$ by number per unit volume of those with Hubble types from Sa to Im. From a sample of 1550 galaxies with $M_\mathrm{B}<-12.5$ in nearby galaxy clusters, Ferguson \& Sandage (\cite{Ferguson & Sandage}) estimate that the galaxies of these Hubble types constitute $\approx 20 \%$ of the total galaxy population (most of the rest being dwarf ellipticals). This would indicate that the local fraction of blue LSBGs is at least $\approx 3.2\%$ by number. Due to the underrepresentation of Sa to Im types in clusters compared to lower-density environments, where many LSBGs may be located (Rosenbaum \& Bomans \cite{Rosenbaum & Bomans}), this fraction should be conservative. 

Assuming $B-V\approx 0.7$ mag (Karick et al. \cite{Karick et al.}) and a stellar population $M/L_V \approx 1.5$ (Mateo et al. \cite{Mateo et al.}) for dE/dSph dwarf galaxies, the Ferguson \& Sandage luminosity limit corresponds to a stellar mass of $\approx 2.4\times 10^7 \ M_\odot$. For the smaller fraction of dI galaxies, an HI mass of $\approx 2 \times 10^7 \ M_\odot$ (Hoffman et al. \cite{Hoffman et al.}) may be inferred at this luminosity. Given typical gas mass fractions for dI galaxies (0.4--0.8; Pilyugin \& Ferrini \cite{Pilyugin & Ferrini}), this converts into stellar population masses not very different from those of dE/dSphs. By taking the risk of overinterpreting small number statistics, one may then from Fig. 4 in Nagamine et al. (\cite{Nagamine et al.}) estimate that only $\approx 0.2$\% of the objects with stellar masses  $2.4\times 10^7 \leq M/M_\odot \leq 10^{10}$ should form at $z<0.5$ (corresponding to an age $<5$ Gyr) and $\approx 0.8\%$ at $z<1.0$ (age $<7.5$ Gyr). The typical redshift of formation for objects of these masses is instead predicted to be $z\approx 3$--5.5 (corresponding to ages of 11--12 Gyr). Actually, the formation redshifts used in Nagamine et al. (\cite{Nagamine et al.}) correspond to the mass-weighted averages of the formation epochs of the stellar particles in their simulation, making this comparison to our absolute age estimates (which are by definition systematically higher) very conservative.

We therefore conclude that if the blue LSBGs in our sample really have ages below 5 Gyr, there would be an order of magnitude (3.2\% compared to 0.2\%) too many such objects in the local universe to be compatible with the Nagamine et al. (\cite{Nagamine et al.}) predictions. If our targets are younger than 7.5 Gyr, the discrepancy is instead a factor of 4 (3.2\% compared to 0.8\%). Hence, the ages of the bluest LSBGs may potentially provide an important test of galaxy formation scenarios. Since LSBGs typically are non-merging disk galaxies with well-behaved rotation curves, they are likely to be superior in this respect compared to other young galaxy candidates, like blue compact and HII-galaxies, for which estimates of both stellar and total mass may be more problematic. Developing methods to constrain the age of the blue LSBG population with higher precision would therefore be very valuable.  

What observations would be required to better constrain the ages? The strength of the 4000 \AA{} break, $D4000$, is often used as an age indicator for stellar populations, but actually depends very weakly on age at low metallicities (e.g. Worthey \cite{Worthey}; Maraston \cite{Maraston}). Hydrogen Balmer lines in absorption appear potentially useful (e.g. Worthey \cite{Worthey}; Vazdekis \& Arimoto \cite{Vazdekis & Arimoto}), but the behaviour of these diagnostics in stellar populations with extended star formation histories is largely unexplored. 

An alternative to using minor spectral features is to use the shape of the continuum over a wider wavelength range. The two degenerate stellar populations discussed in Sect. 4.2 (a 13 Gyr old, $\tau=15$ Gyr population and a 3 Gyr old, $\tau=1$ Gyr population) have nearly identical continua throughout the optical and near-IR part of the spectrum. As one moves to shorter wavelengths, this is however no longer the case. In the mid- to far-ultraviolet range (1000--3000 \AA), the slopes of the continua for these two populations differ significantly. Hence, the addition of UV data could substantially improve the situation. Even the inclusion of $U$-band photometry would be useful.

In this paper, EW(H$\alpha$) has admittedly been used in a very crude way, mainly due to the difficulties involved in sensibly weighting emission line data against photometry. A more intelligent age fitting procedure could possibly improve the constraints. The effects of underlying absorption and selective extinction (Calzetti et al. \cite{Calzetti et al.}) would then however also have to be more carefully considered.

Even if the ages turned out to be very low, it would be premature to claim a contradiction with the underlying $\Lambda$CDM cosmology, since current simulations still have difficulties in forming realistic disk galaxies (e.g. Primack \cite{Primack}). A particular concern is that there may exist mechanisms which delay star formation, so that while mass assembly of LSBGs takes places early in the history of the universe, in accord with $\Lambda$CDM, significant star formation does not occur until much later. Photoionization by the extragalactic UV background has been advocated as one way to suppress star formation in very low-mass galaxies (Dong et al. \cite{Dong et al.}), but would not be efficient for objects in the range of total masses relevant for our galaxies (probably $\sim 10^{8}$--$10^{11} \ M_\odot$ after inclusion of dark matter). A more viable mechanism is perhaps suppressed star formation due to high angular momentum. One popular explanation for the low star formation efficiency in LSBG is that these galaxies formed inside dark matter halos with unusually high spin parameters (e.g. Dalcanton et al.\cite{Dalcanton et al.}; Jimenez et al. \cite{jimenez1}; Boissier et al. \cite{Boissier et al.}). A similar scheme has also been suggested to explain the extremely low star formation rates  in the hypothetical dark galaxy population (e.g. Jimenez et al. \cite{jimenez2}). Until processes like these, which could substantially affect the SFH of stellar populations, have been considered in the predictions of the average formation redshifts for the stars in galaxies, the assumed cosmology cannot really be blamed for any potential age discrepancy of the type discussed here.

\section{Summary}
In a critical assessment of the possibility to derive properties of the bluest low surface brightness galaxies from optical/near-IR broadband photometry and H$\alpha$ emission line data, we find that:

\begin{itemize}
\item The global EW(H$\alpha$) observed in the bluest LSBGs are too low to be consistent with recently proposed scenarios involving constant or increasing star formation rates over cosmological time scales (Boissier et al. \cite{Boissier et al.}), unless the fraction of high-mass stars is significantly lower than predicted by the Salpeter IMF with $M_\mathrm{up}=120 \ M_\odot$. A combination of EW(H$\alpha$) and chemical abundances may provide additional constraints on both the star formation history and IMF, and should be investigated further.
\item The ages of these objects are poorly constrained without prior knowledge of the star formation history. In fact, current observations cannot rule out the possibility that these objects formed as recently as 1--2 Gyr ago. Additional UV observations may however improve the age estimates.
\item If low ages are typical for the bluest LSBGs, this would contradict current models of hierarchical galaxy formation in the $\Lambda$CDM cosmology, since only a small number of galaxies with stellar masses typical for our objects are predicted to form this late in the history of the universe. 
\item Contrary to common methodology when estimating the masses of stellar populations, we note that the uncertainties associated with stellar $M/L$ are not necessarily smaller in the near-IR filters. Instead, the $I$-band appears to be optimal.
\end{itemize}

\begin{acknowledgements} This work was partly supported by the
Swedish Natural Science Research Council. Our referee, M. Fioc, is acknowledged for very detailed and helpful comments on the manuscript.
\end{acknowledgements}

\end{document}